\documentclass[aps,prd,preprint,superscriptaddress]{revtex4}
\usepackage{graphicx}
\usepackage{epstopdf}
\usepackage{amsmath,amssymb,bm,graphics}

\usepackage[usenames]{color}
\usepackage[colorlinks=true, urlcolor=navyblue, linkcolor=navyblue, citecolor=navyblue]{hyperref}
\usepackage{relsize}
\definecolor{navyblue}{rgb}{0,0.08,0.45}

\usepackage{color}

\newcommand{\beq}{\begin{equation}}
\newcommand{\enq}{\end{equation}}
\newcommand{\beqa}{\begin{eqnarray}}
\newcommand{\beqast}{\begin{eqnarray*}}
\newcommand{\enqa}{\end{eqnarray}}
\newcommand{\enqast}{\end{eqnarray*}}
\newcommand{\beml}{\begin{multline}}
\newcommand{\enml}{\end{multline}}
\newcommand{\nn}{\nonumber}
\newcommand{\req}[1]{(\ref{#1})}
\newcommand{\lb}{\label}

\newcommand{\cM}{\mathcal{M}}

\newcommand{\pa}{\partial}

\newcommand{\bec}{\begin{center}}
\newcommand{\enc}{\end{center}}
\newcommand{\beqo}{\begin{quote}}
\newcommand{\enqo}{\end{quote}}

\newcommand{\half}{{\textstyle{\frac{1}{2}}}}
\newcommand{\mbf}[1]{\mathbf{#1}}

\newcommand{\al}{\alpha}

\newcommand{\ga}{\gamma}
\newcommand{\de}{\delta}
\newcommand{\ep}{\epsilon}

\newcommand{\et}{\eta}

\newcommand{\la}{\lambda}

\newcommand{\vp}{\varphi}
\newcommand{\vph}{\varphi}

\newcommand{\om}{\omega}

\newcommand{\Ga}{\Gamma}

\newcommand{\La}{\Lambda}

\newcommand{\Si}{\Sigma}

\newcommand{\Ph}{\Phi}

\newcommand{\Om}{\Omega}

\begin{document}

\preprint{SLAC-PUB-15335}

\title{Kinematical and Dynamical Aspects of Higher-Spin Bound-State Equations in Holographic QCD}

\author{Guy F. de T\'eramond}
\email[]{gdt@asterix.crnet.cr}
%\thanks{}
\affiliation{Universidad de Costa Rica, San Jos\'e, Costa Rica}

\author{Hans G\"unter Dosch}
\email[]{h.g.dosch@thphys.uni-heidelberg.de}
%\thanks{}
\affiliation{Institut f\"ur Theoretische Physik, Philosophenweg
16, D-6900 Heidelberg, Germany}

\author{Stanley J. Brodsky}
\email[]{sjbth@slac.stanford.edu}
%\thanks{}
\affiliation{SLAC National Accelerator Laboratory, Stanford University,
Stanford, California 94309, USA}

\date{\today}

\begin{abstract}

In this paper we derive holographic wave equations for hadrons with arbitrary spin starting from an effective action in a higher-dimensional space asymptotic to   anti-de Sitter (AdS) space.   Our procedure takes advantage of the local tangent frame, and it applies to all spins, including half-integer spins.  An essential element is the mapping of the higher-dimensional equations of motion to the light-front Hamiltonian,  thus allowing a clear distinction between the kinematical and dynamical aspects of the holographic approach to hadron physics.  Accordingly, the non-trivial geometry of pure AdS space encodes the kinematics,  and the additional deformations of AdS space encode the dynamics, including confinement.  It thus becomes possible to identify the features of holographic QCD which are independent of the specific mechanisms of conformal symmetry breaking.   In particular, we account for some aspects of the striking similarities and differences observed in the systematics of the meson and baryon spectra.

\end{abstract}

%\pacs{11.15.Bt, 12.20.Ds}
%\keywords{}

\maketitle

\section{Introduction}

Quantum Chromodynamics provides a description of hadrons in terms of fundamental quark and gluon fields appearing in the QCD Lagrangian.
Because of its strong coupling nature, calculations of hadronic properties, such as hadron masses and color-confinement, still remain among the most challenging dynamical problems in hadron physics.
Euclidean lattice methods~\cite{Wilson:1974sk} provide an important first-principles numerical simulation of nonperturbative QCD.   However, the excitation spectrum of hadrons represents an important challenge to lattice QCD due to the enormous computational complexity beyond ground-state configurations and the unavoidable presence of multi-hadron thresholds. 
Furthermore, dynamical observables in Minkowski space-time are not  obtained directly from Euclidean space lattice computations.   Dyson-Schwinger and Bethe-Salpeter methods have also led to many important insights, such as the infrared fixed-point behavior of the  strong coupling constant and the behavior of the quark running mass~\cite{Cornwall:1981zr}. However, in practice, these analyses have been limited to ladder approximation in Landau gauge~\cite{Maris:2003vk}.

The AdS/CFT correspondence between gravity on a higher-dimensional anti--de Sitter (AdS) space and conformal field theories (CFT) in physical space-time~\cite{Maldacena:1997re},
has led to a semiclassical approximation for strongly-coupled quantum field theories
which provides physical insights into its nonperturbative dynamics. The correspondence is holographic in the sense that it determines a duality between  theories in different number of space-time dimensions.  In practice, the duality provides an effective gravity description in a ($d+1$)-dimensional AdS
space-time in terms of a flat $d$-dimensional conformally-invariant quantum field theory defined at the AdS
asymptotic boundary~\cite{Gubser:1998bc, Witten:1998qj}.  As we discuss below, the equations of motion in AdS space have a remarkable holographic mapping to the equations of motion obtained in light-front Hamiltonian theory~\cite{Dirac:1949cp} (Dirac's front form) in physical space-time. 
Thus, in principle, one can compute physical observables in a strongly coupled gauge theory  in terms of an effective classical gravity theory.

Anti-de Sitter AdS$_{d+1}$ space is a maximally symmetric space-time
with negative curvature and a $d$-dimensional space-time boundary.
The most general group of transformations that leave invariant the AdS$_{d+1}$ differential line element
\begin{equation} \label{AdSz}
ds^2 = \frac{R^2}{z^2} \left( dx_\mu dx^\mu - dz^2\right),
\end{equation}
 the isometry group, has  $(d+1)(d+2)/2$ dimensions ($R$ is the AdS radius). Five-dimensional anti-de Sitter space AdS$_5$ has 15 isometries, in correspondence with the number of generators of the conformal group in four dimensions.
Since the AdS metric (\ref{AdSz}) is invariant under a dilatation of all coordinates $x^\mu  \to \lambda x^\mu$  and $z \to \lambda z$, it follows that the additional dimension, with holographic variable $z$, acts like a scaling variable in Minkowski space: different values of $z$ correspond to
different energy scales at which the hadron is examined.  As a result,
a short space-like or time-like invariant interval
near the light-cone, $x_\mu x^\mu \to 0$ maps to the conformal  AdS boundary near $z \to 0$.  On the other hand, a large invariant four-dimensional  interval of
confinement dimensions $x_\mu x^\mu \sim 1/\Lambda_{\rm QCD}^2$ maps to the large infrared (IR) region
of AdS space $z \sim 1 / \Lambda_{\rm QCD}$.

QCD is fundamentally different from conformal theories since its scale invariance is broken  by quantum effects.
A precise gravity dual to QCD is not known, but the mechanisms of
confinement can be incorporated in the gauge/gravity
correspondence by breaking the maximal symmetry of AdS space, thus inducing a breaking of the conformal symmetry of QCD in  four-dimensional 
space-time. This breaking is effective in the  large infrared (IR) domain  of AdS, $z \sim 1/\Lambda_{\rm QCD}$, and
sets the scale of the strong
interactions~\cite{Polchinski:2001tt}.  In this simplified
approach, the propagation of hadronic modes can be analyzed in a fixed
effective gravitational background asymptotic to AdS space which
encodes essential properties of the QCD dual theory, such as the ultraviolet
(UV) conformal limit from the AdS boundary, as
well as effective modifications of the AdS background geometry in the large-$z$
IR region.  Since the conformal behavior is retained at $z \to 0$,   the modified theory  generates the point-like hard behavior
expected from QCD~\cite{Brodsky:1973kr, Matveev:ra}, instead of the soft behavior characteristic of
extended objects~\cite{Polchinski:2001tt}.

Since AdS space has maximal symmetry, it is a space with constant curvature and does not lead to confinement. 
One possible way to introduce an effective confinement potential  is a sharp cut-off in the infrared region of AdS space, as in the ``hard-wall'' model~\cite{Polchinski:2001tt}, 
where one considers  a slice  of AdS space, $0 \leq z \leq z_0$, and
imposes boundary conditions on the fields at the IR border $z=z_0$.  One can also  use a ``dilaton" background in the holographic coordinate to produce a smooth cutoff at large distances  as  in the ``soft-wall'' model~\cite{Karch:2006pv} which explicitly breaks the maximal  AdS symmetry; this introduces an effective $z$-dependent curvature in the infrared which leads to conformal symmetry breaking in QCD.  Furthermore, one can impose from the onset a correct phenomenological confining structure to determine the effective IR warping of AdS space, for example, by adjusting the dilaton background to reproduce the observed linear Regge behavior of the hadronic mass spectrum $M^2$ as a function of the excitation quantum numbers~\cite{Karch:2006pv, Shifman:2005zn}. A convenient feature of 
the approach described below is that the dilaton background can be absorbed into a universal (spin-independent) warp of the AdS metric. One can also consider models where the dilaton field is dynamically coupled to gravity~\cite{Csaki:2006ji, Gursoy:2007er}.

Hadronic  states in AdS space are represented by modes $\Phi_P(x, z) = e^{ i P \cdot x} \Phi(z) \ep(P)$, with plane
waves along Minkowski coordinates $x^\mu$ and a  normalizable profile function
$\Phi(z)$ along the holographic coordinate $z$. The hadronic invariant mass states $P_\mu P^\mu = M^2$  are found by solving
the eigenvalue problem for the AdS wave equation. The spin degrees
of freedom are encoded in the tensor or generalized Rarita-Schwinger spinor $\ep(P)$. A physical hadron has
polarization indices  along the $d$ physical coordinates, all other components vanish identically.

Light-front (LF)  holographic methods  were originally
introduced~\cite{Brodsky:2006uqa} by matching the  electromagnetic
current matrix elements in AdS space~\cite{Polchinski:2002jw} with
the corresponding expression  derived from light-front  quantization in
physical space time. It was also shown that one obtains  identical
holographic mapping using the matrix elements of the
energy-momentum tensor~\cite{Brodsky:2008pf} by perturbing the AdS
metric (\ref{AdSz}) around its static solution~\cite{Abidin:2008ku}, thus establishing a precise relation between wave functions in AdS space and the light-front wavefunctions describing the internal structure of hadrons.

Unlike ordinary instant-time quantization, light-front Hamiltonian equations of motion are frame independent; remarkably they  have a structure which matches exactly the eigenmode equations in AdS space. This makes a direct connection of QCD with AdS methods possible.  In fact, one can
derive the light-front holographic duality of AdS  by starting from the light-front Hamiltonian equations of motion for a relativistic bound-state system
in physical space-time~\cite{deTeramond:2008ht}.
To a first semiclassical approximation, where quantum loops and quark masses
are not included, this leads to a LF Hamiltonian equation which
describes the bound state dynamics of light hadrons  in terms of
an invariant impact variable $\zeta$, which measures the
separation of the partons within the hadron at  fixed light-front time $\tau = t+z/c$~\cite{Dirac:1949cp}. This allows one to identify the  variable $z$ in
AdS space with the impact variable $\zeta$~\cite{Brodsky:2006uqa,
Brodsky:2008pf, deTeramond:2008ht},  thus giving  the holographic
variable a precise definition and very intuitive meaning in light-front QCD.

Remarkably, the pure AdS  equations correspond to the light-front kinetic energy of  the partons
inside a hadron, whereas the light-front interactions  which build confinement
correspond to the truncation of AdS space in an effective dual
gravity approximation~\cite{deTeramond:2008ht}. From this point of view, the non-trivial geometry of pure AdS space encodes  the kinematical aspects and additional
deformations of AdS space  encode dynamics, including confinement. For example, in the hard-wall model, dynamical aspects are implemented by boundary conditions on the hadronic eigenmodes.  The  geometry of AdS space then leads to terms in the equation of motion which are identified with the orbital angular momentum of the constituents in light-front quantized QCD. This identification  is a key element in the description of the internal structure of hadrons using LF holographic principles.

The  treatment of higher-spin states in the ``bottom-up"
approach to holographic QCD described above is an important
touchstone for this procedure. Up to now there are essentially two
systematic bottom-up approaches to describe higher-spin hadronic modes in holographic QCD: one by Karch, Katz, Son, and Stephanov
(KKSS)~\cite{Karch:2006pv}, which is based in the usual AdS/QCD framework where background
fields are introduced to match the chiral symmetries of QCD~\cite{Erlich:2005qh, DaRold:2005zs}, but
without explicit connection with the internal constituent
structure of hadrons~\cite{Brodsky:2003px}; and the other  by two of the
authors~\cite{Brodsky:2006uqa, Brodsky:2008pf, deTeramond:2008ht}, 
using as a starting  point the precise mapping of  AdS equations
to gauge theories quantized on the light-front, as discussed above. Various
other approaches  follow more or less these lines~\cite{BoschiFilho:2005yh, dePaula:2008fp, Gherghetta:2009ac, Gutsche:2011vb}.

The description of higher-spin modes in AdS space is a notoriously
difficult problem~\cite{Fronsdal:1978vb, Fradkin:1986qy, Buchbinder:2001bs, Metsaev:2003cu}, and thus  there is much interest in finding a simplified
approach which can describe higher-spin hadrons using the  gauge/gravity
duality. For example,  the approach of~\cite{ deTeramond:2008ht}
relies on rescaling the solution of a scalar field $\Phi(z)$ by
shifting dimensions introducing a spin dependent factor~\cite{deTeramond:2008ht,
deTeramond:2012rt}. This procedure  is based on  the conformal
structure of AdS/CFT and the close relation between  AdS/CFT and the light-front
approach~\cite{deTeramond:2008ht}.

The KKSS approach~\cite{Karch:2006pv}   starts from a gauge invariant
action in  AdS space, and uses
the gauge invariance of the model to construct explicitly an effective action in
terms  of higher-spin modes with only the physical degrees of freedom. However, this
approach is not applicable to pseudoscalar particles and their trajectories, and their angular
excitations do not lead to a relation with light-front quantized QCD, which is an essential point of the approach
described in Ref~\cite{deTeramond:2008ht}.

In this paper we start from a manifestly covariant effective action
constructed with AdS tensors or generalized Rarita-Schwinger spinor
fields in AdS space for all integer and half-integer spins, respectively. The occurrence of covariant derivatives with affine connections complicates the
Euler-Lagrange equations for the various actions that are considered, but it will be
shown that the transition to the Lorentz frame (the local frame with tangent indices)
simplifies matters considerably.  Further simplification is brought by the fact that physical hadrons have tensor indices along the 3 + 1 physical
coordinates and by the precise mapping of the AdS equations to the light-front equations of motion at equal light-front time, thus providing a clear distinction between the kinematical and dynamical aspects of the problem.

The derivation of the Euler-Lagrange equations of motion for higher integer and half-integer spin is in general severely complicated by the constraints imposed by the subsidiary conditions necessary to eliminate the lower spin states from the symmetric tensors and Rarita-Schwinger spinors~\cite{SW1:1958}. In our approach these subsidiary conditions follow from the general covariance of the higher dimensional effective action. 
We then can systematically treat the resulting different
approaches to conformal symmetry breaking and the consequences for the
hadron spectrum. In particular, we will give a systematic derivation of
the  phenomenologically successful approach given in~\cite{deTeramond:2008ht} which leads to a massless pion in the chiral limit, and linear Regge trajectories with the same slope in orbital angular momentum $L$ and node number $n$~\cite{deTeramond:2012rt}.

This paper is organized as follows: we discuss the equations of motion  for
general  integer spin in a higher-dimensional background in Sec. \ref{int}  and the  corresponding holographic mapping to the light-front Hamiltonian equations 
in Sec. \ref{LFmapM}. The wave equations for higher half-integer spin is described in Sec. \ref{half} and their mapping to light-front physics in Sec.  \ref{LFmapB}. We summarize and discuss the final results 
in Sec. \ref{summ}.  Technical details of the calculations are collected in  Appendix \ref{intspinA} for integer spin and in Appendix \ref{intspinB} for half-integer spin.

\section{Integer Spin \label{int}}

We will begin with the  formulation of bound-state equations for mesons of arbitrary spin $J$ in a higher-dimensional  AdS space. As we shall show below, there is a remarkable correspondence between the equations of motion in AdS space and the Hamiltonian equation for the relativistic bound-state system for the corresponding angular momentum in light-front theory.

\subsection{Invariant Action and Equations of Motion \label{integer}}

The coordinates of AdS$_{d+1}$ space are the $d$-dimensional Minkowski coordinates
$x^\mu$ and the holographic variable $z$. The combined coordinates are  labeled $x^M =
\left(x^\mu, z\right)$, with $M, N = 0, \dots , d$ the indices of the higher dimensional $d+1$ curved space, and
 $\mu, \nu= 0, 1, \dots, d-1$ the Minkowski flat space-time indices. In Poincar\'e coordinates, $z \ge 0$, the conformal AdS metric is
\beqa \label{AdSmetric} \nn
ds^2 &=&  g_{M N} dx^M dx^N \\
&=& \frac{R^2}{z^2}  \left( \eta_{\mu \nu} dx^\mu
dx^\nu - dz^2\right), 
\enqa 
and thus the metric tensor $g_{MN}$ 
\beq
g_{MN} = \frac{R^2}{z^2} \eta_{MN},  \hspace{10pt} g^{MN} = \frac{z^2}{R^2} \eta^{MN},
\label{metricAdS}
 \enq
 where $\eta_{M N}$ is the flat $d + 1$ Minkowski metric  $(1, -1, \cdots, - 1)$.

Fields with integer spin  in AdS$_{d+1}$ are represented  by a rank-$J$ tensor field
 $\Phi(x^M)_{N_1 N_2 \dots N_J}$ which is totally symmetric in all its indices. 
 Such a tensor contains  lower spins, which can be eliminated by imposing the subsidiary conditions defined below.
The action for a spin-$J$ field in AdS$_{d+1}$ space-time in the presence of a dilaton background field $\varphi(z)$  is given by 
\begin{multline}
\label{action1}
 S = \int d^{d} x \,dz \,\sqrt{\vert g \vert}  \; e^{\vp(z)} \,g^{N_1 N_1'} \cdots  g^{N_J N_J'}   \Big(  g^{M M'} D_M \Phi^*_{N_1 \dots N_J}\, D_{M'} \Phi_{N_1 ' \dots N_J'}  \\
 - \mu^2  \, \Phi^*_{N_1 \dots N_J} \, \Phi_{N_1 ' \dots N_J'} + \cdots \Big),
 \end{multline}
where  $\sqrt{\vert g \vert} = (R/z)^{d+1}$ and $D_M$ is the covariant derivative which includes the affine connection (Appendix \ref{der-int}).
At this point, the  higher  dimensional mass $\mu$ in  (\ref{action1}) is not a physical observable and is {\it a priori} an arbitrary
parameter. The omitted terms  in the action, indicated by $\cdots$,  refer to terms with different contractions. 
The dilaton background $\varphi(z)$ in  (\ref{action1})  introduces an energy scale in the  AdS action, thus breaking conformal invariance. 
It is a function of the holographic coordinate $z$, and it is assumed to vanish
in the conformal ultraviolet limit $z \to 0$.

Inserting the covariant derivatives in the  action leads to a rather complicated expression. Furthermore,
for higher-spin actions, the additional terms from different contractions in (\ref{action1}) bring an enormous complexity. 
A physical hadron has  polarization indices along the 3~+~1 physical coordinates, $\Phi_{\nu_1 \nu_2 \cdots \nu_J}$. All other components  must vanish identically
\beq \label{no}
\Phi_{z N_2 \cdots N_J}  = 0.
\enq
This brings a considerable simplification in (\ref{action1}) since 
we only have to consider the  subspace of tensors which are orthogonal to
the holographic dimension.  As we shall see, the constraints imposed by the mapping of the AdS equations of motion to the light-front Hamiltonian in physical space-time for the hadronic bound-state system at fixed LF time will give us further insight since it allows an explicit distinction between kinematical and dynamical aspects.

As a practical procedure, we will construct an effective action with  a $z$-dependent effective AdS mass $\mu_{\it eff}(z)$ in the action, which can
absorb  the contribution from different contractions in (\ref{action1}).  Our effective action $S_{\it eff}$ is
\begin{multline}
\label{action2}
S_{\it eff} = \int d^{d} x \,dz \,\sqrt{\vert g \vert}  \; e^{\vp(z)} \,g^{N_1 N_1'} \cdots  g^{N_J N_J'}   \Big(  g^{M M'} D_M \Phi^*_{N_1 \dots N_J}\, D_{M'} \Phi_{N_1 ' \dots N_J'}  \\
 - \mu_{\it eff}^2(z)  \, \Phi^*_{N_1 \dots N_J} \, \Phi_{N_1 ' \dots N_J'} \Big),
 \end{multline}
where the  function $\mu_{\it eff}(z)$, which encodes kinematical aspects of the problem, is {\it a priori} unknown.  But, as we shall show below, the additional symmetry breaking due to the $z$-dependence of the effective mass allows a clear separation of kinematical and dynamical effects. In fact, the $z$ dependence can be determined either by the precise mapping of AdS to light-front physics, or by eliminating interference terms between kinematical and dynamical effects. The agreement between the two methods shows how the light-front mapping and the explicit separation of kinematical and dynamical effects are intertwined.

The equations of motion are  obtained from the Euler-Lagrange 
equations in the subspace defined by (\ref{no})
 \beq \label{ELJ} 
 \frac{ \de S_{\it eff}}{ \de \Phi^*_{\nu_1 \nu_2 \cdots \nu_J}} = 0,
\enq
 and
 \beq \label{ELz}
  \frac{\de S_{\it eff}}{\de \Ph^*_{ z N_2 \cdots N_J}} = 0.
 \enq
The wave equations for hadronic modes follow from the Euler-Lagrange equation (\ref{ELJ}) for tensors orthogonal to the holographic coordinate $z$. But  remarkably, as we will show below, terms in the action which are linear in tensor fields, with one or more indices along the holographic direction, $\Phi_{z N_2 \cdots N_J}$, give us from (\ref{ELz})
 the kinematical constraints required to eliminate the lower-spin states.

The covariant derivatives $D_M$ are given in Appendix \ref{intspinA}.  As shown  there, it is useful to introduce 
fields with tangent indices using a local Lorentz frame, the inertial frame
\beq \label{VB}
\hat \Phi_{A_1 A_2 \cdots A_J}
 = e_{A_1}^{N_1} e_{A_2}^{N_2} \cdots e_{A_J}^{N_J} \,
 \Phi_{N_1 N_2 \cdots N_J},
 \enq
 where the vielbein $e^A_M$ is obtained from a transformation to a local tangent frame, $g_{M N} = e^A_M e^B_N  \eta_{A B}$, and the indices 
  $A, B = 0, \dots , d$  are the indices in the  space tangent to AdS$_{d+1}$.
 The local tangent metric  $\eta_{A B}$ has diagonal components  $(1, -1, \cdots, -1)$.   
 In AdS space 
 \beq  \label{VBAdS}
 e^A_M = \frac{R}{z} \delta^A_M, ~~~    e^M_A = \frac{z}{R} \delta^M_A,
 \enq
and thus
 \beq \label{inertial}
  \hat  \Phi_{N_1 \dots N_J}=\left(\frac{z}{R}\right)^J\,
 \,  \Phi_{N_1 \dots N_J}.
\enq
Notably,  one can express the covariant derivatives in a general frame in terms of partial derivatives in a local tangent frame. We find
\beq \label{dzm}
  D_z \Phi_{N_1 \dots N_J}= \left(\frac{R}{z}\right)^J \pa_z \hat
\Phi_{N_1 \dots N_J},
 \enq
 and
 \begin{multline} \label{dmum}  
 \lefteqn{g^{\mu \mu'}  g^{\nu_1 \nu_1'}\dots  g^{\nu_J \nu_J'}
D_\mu \Phi_{\nu_1 \dots \nu_J} \, D_{\mu'} \Phi_{\nu_1' \dots \nu'_J}= } \\
 g^{\mu \mu'}  \eta^{\nu_1 \nu_1'}\dots  \eta^{\nu_J \nu_J'} 
 \left( \pa_\mu \hat \Phi_{\nu_1 \dots \nu_J} \,
\pa_{\mu'} \hat \Phi_{\nu_1' \dots \nu'_J} 
+  g^{zz}   J \, \Omega^2(z) \, \hat \Phi_{\nu_1 \dots \nu_J}\, \hat
\Phi_{\nu_1' \dots \nu'_J}\right) ,
\end{multline}
where $\Omega(z)= 1/z$ is the  AdS warp factor in the affine connection as shown in Appendix~\ref{der-int}.

We split the action ({\ref{action2}) into three terms, a term $S^{[0]}_{\it eff}$ which
contains only fields $\Phi_{\nu_1 \dots \nu_J}$
orthogonal to the holographic direction, and a term $S^{[1]}_{\it eff}$, which
is linear in the fields  $\Phi^*_{z N_2 \cdots N_J}$,   $\Phi^*_{N_1 z \cdots N_J}$, $\cdots$,   $\Phi^*_{N_1  N_2 \cdots z}$.
The remainder  is quadratic in fields with $z$-components, {\it i.~e.}, it contains terms such as  $\Phi^*_{z N_2 \dots N_J}\Phi_{z N_2' \dots N_J'}$. This last term
 does not contribute to the Euler-Lagrange equations \req{ELz}, since upon variation of the action, a vanishing term (\ref{no}) is left.

 Using (\ref{action2}), \req{dzm} and \req{dmum} we find 
\begin{multline}
 \label{a-wraph}
S^{[0]}_{\it eff} = \int d^{d}x\, dz\, \left(\frac{R}{z}\right)^{d-1}
 e^{\vp(z)}\,  \eta^{\nu_1  \nu _1'} \cdots  \eta^{\nu_J  \nu _J'}
 \Bigg( - \pa_z \hat \Phi^*_{\nu_1 \dots \nu_J}\,\pa_z \hat \Phi_{\nu_1' \dots \nu_J'}   \\
   + \eta^{\mu \mu'}
\pa _\mu \hat \Phi^*_{\nu_1 \dots \nu_J}\,\pa_{\mu'} \hat \Phi_{\nu_1' \dots \nu_J'} 
 - \left[ \left(\frac{\mu_{\it eff} (z) R}{z}\right)^2  + J \, \Omega^2(z)\right]
\hat \Phi^*_{\nu_ 1 \dots \nu_J}\, \hat \Phi_{\nu_1' \dots \nu_J'}\Bigg),
\end{multline}
 and
  \begin{multline}
\label{a-subh}
 S_{\it eff}^{[1]} = \int  d^{d}x\, dz\, \left(\frac{R }{z}\right)^{d-1} 
  e^{\vp(z)} \,
\Big( - J \, \Omega(z) \,  \eta^{\mu \mu'}
\eta^{N_2 \nu \,_2'} \cdots \eta^{N_J \nu _J'}
  \pa_\mu \hat \Phi^*_{z N_2 \dots N_J} \hat  \Phi_{\mu' \nu _2' \dots \nu_J'}\\
+ J  \, \Omega(z) \,  \eta^{\mu \nu} \eta^{N_2 \nu \,_2'} \cdots \eta^{N_J \nu_J'}
 \hat \Phi^*_{z N_2 \dots N_J}
 \pa_{\mu} \hat \Phi_{\nu \nu _2' \dots \nu_J'}  \\
 -J(J-1)  \,  \Omega^2(z) \, 
\eta^{\mu \nu}
 \eta^{N_3 \nu \,_3'} \cdots \eta^{N_J \nu_J'}
\hat \Phi^*_{z z N_3 \cdots N_J} \hat \Phi_{\mu \nu \nu_3'
\cdots \nu_J'} \Big) .
 \end{multline}
As can be seen from the presence of the affine warp factor $\Omega(z)$ in (\ref{a-subh}), this last term is only due to the affine connection  and thus should only contribute to  kinematical constraints.

From \req{a-wraph}  we obtain, upon variation with respect to $ \hat \Phi^*_{\nu_1 \dots \nu_J}$ (\ref{ELJ}),
 the equation of motion in the local tangent space
 \beq  \label{PhiJhat}
 \left[  \pa_\mu \pa^\mu 
   -  \frac{ z^{d-1}}{e^{\varphi(z)}}   \partial_z \left(\frac{e^{\varphi(z)}}{z^{d-1}} \partial_z   \right) 
  +  \frac{(\mu_{\it eff} (z) R)^2 + J }{z^2}  \right]  \hat \Phi_{\nu_1 \dots \nu_J} = 0,
  \enq
where $\pa_\mu \pa^\mu \equiv \eta^{\mu \nu} \pa_\mu \pa_\nu$.  

From (\ref{PhiJhat}) and (\ref{inertial}) we can now write the wave equation in a general frame
in terms of the original covariant tensor field $\Phi_{N_1 \cdots N_J}$
\beq  \label{PhiJ}
 \left[  \pa_\mu \pa^\mu 
   -  \frac{ z^{d-1 - 2J}}{e^{\varphi(z)}}   \partial_z \left(\frac{e^{\varphi(z)}}{z^{d-1 - 2 J}} \partial_z   \right) 
  +  \frac{(mR)^2 }{z^2}  \right]   \Phi_{\nu_1 \dots \nu_J} = 0,
  \enq
with  
  \beq \label{muphi} 
  (m\, R)^2 =(\mu_{\it eff}(z) R)^2  - J z \, \vp'(z) + J(d - J +1) ,
  \enq
  which is  the result found in Refs.~\cite{deTeramond:2008ht, deTeramond:2012rt} by rescaling the wave equation for a scalar field.

From  \req{a-subh} we obtain by variation with respect to
$ \hat \Phi^*_{N_1 \cdots z  \cdots N_J}$  (\ref{ELz}) the kinematical constraints which  eliminate lower spin states from the symmetric field tensor
\beq \label{scPhi}
\eta^{\mu \nu} \pa_\mu  \Phi_{\nu \nu_2 \cdots \nu_J}=0, \quad
\eta^{\mu \nu}   \Phi_{ \mu \nu \nu_3  \cdots \nu_J}=0.
\enq
It is remarkable that we have started in AdS space with unconstrained
symmetric spinors, but the non-trivial affine connection of AdS geometry gives us precisely the subsidiary
conditions to eliminate the lower spin states $J-1, \,J-2, \cdots$ from the fully symmetric tensor field.  We note
that the conditions \req{scPhi} are independent of the conformal symmetry breaking terms in the action, since they are a consequence of the kinematical aspects encoded in the AdS metric.

A free hadronic state in holographic QCD  is described by a plane wave in physical space-time, a  $z$-independent spinor $\ep_{\nu_1 \cdots \nu_J}$ with polarization indices along physical coordinates and a $z$-dependent profile function:
\beq \label{scalarcov}
\Phi_{\nu_1 \cdots \nu_J}(x, z) = e ^{ i P \cdot x} \,  \Phi_J(z) \,  \ep_{\nu_1 \cdots \nu_J}({P}),
\enq 
with invariant hadron mass $P_\mu P^\mu \equiv \eta^{\mu \nu} P_\mu P_\nu = M^2$.  Inserting  (\ref{scalarcov}) into the wave equation (\ref{PhiJ}) we obtain the bound-state
eigenvale equation 
\beq  \label{PhiJM}
 \left[ 
   -  \frac{ z^{d-1- 2J}}{e^{\varphi(z)}}   \partial_z \left(\frac{e^{\varphi(z)}}{z^{d-1-2J}} \partial_z   \right) 
  +  \frac{(m\,R )^2}{z^2}  \right]  \Phi_J = M^2 \Phi_J,
  \enq
  where the  normalizable solution  $ \Phi_J$  from  the eigenvalue equation (\ref{PhiJM}) is normalized according to
\beq  \label{Phinorm}
R^{d - 1 - 2 J} \int_0^{\infty} \! \frac{dz}{z^{d -1 - 2 J}} \, e^{\varphi(z)} \Phi_J^2 (z)  =
R^{d - 1} \int_0^{\infty} \! \frac{dz}{z^{d -1}} \, e^{\varphi(z)} \hat \Phi_J^2 (z) =1.
\enq
 We also recover from (\ref{scPhi}) and (\ref{scalarcov}) the kinematical constraints
\beq \label{sub-spin}
 \eta^{\mu \nu } P_\mu \,\ep_{\nu \nu_2 \cdots \nu_J}=0, \quad
\eta^{\mu \nu } \,\ep_{\mu \nu \nu_3  \cdots \nu_J}=0.
 \enq

In the case of a scalar field, the covariant derivative is  the usual partial derivative, and there are no additional contractions in the action;  thus $\mu_{\it eff} = \mu = m$ is a constant.  
For a spin-1 wave equation, there is one additional term from the antisymmetric contraction, and the contribution from the parallel transport cancels out.  It is also simple in this case to determine the effective mass $\mu_{\it eff}$ in (\ref{action2}) by the comparison with the full expression for the action of a vector field (which includes the antisymmetric contraction). This is shown in the Appendix \ref{vect}. Thus for spin-1, we have $\mu = m$ and  $(\mu_{\it eff}(z) R)^2 = ( \mu R)^2   +  z \, \vp'(z) - d$.

In general, the AdS mass $m$ in the wave equation (\ref{PhiJ})  or (\ref{PhiJM}) is determined from the mapping  to the light-front Hamiltonian, as we will show in the next section.  Since  $m$ will map  to the Casimir operator of the orbital angular momentum in the light-front (a kinematical quantity) it follows that $m$  should be a constant.  Consequently, the $z$-dependence of the effective mass (\ref{muphi})
\beq \label{muphiz} 
 (\mu_{\it eff}(z) R)^2 = (mR)^2 + J z \, \vp'(z)  - J(d - J + 1) ,
 \enq
in the AdS action (\ref{action2}) is determined {\it a posteriori} by kinematical constraints in the light-front,  namely that the mass $m$ in \req{PhiJ} or or (\ref{PhiJM}) must be a constant.

Our demand that the kinematical  and dynamical  effects are clearly separated in the equations of motion gives us 
a complementary way to arrive to the  $z$-dependence of  the effective mass $\mu_{\it eff}(z)$ (\ref{muphiz}).  In general,
the presence of a dilaton in the effective action (\ref{action2}) and the quadratic appearance of covariant derivatives leads  to a mixture of kinematical and dynamical effects. But, as is shown in the Appendix \ref{app-el}, an appropriate $z$ dependence of the effective mass term can cancel these interference terms. This requirement determines the $z$ dependence completely and leads again to the relation \req{muphiz}.

\subsubsection{Confining Interaction and Warped Metrics}

In the Einstein frame the dilaton term is absent and the maximal symmetry of AdS space is broken by the introduction of an additional $J$-independent warp factor 
in the AdS metric in order to include confinement forces
\beqa  \label{gw}
ds^2 &=& \tilde g_{M N} dx^M dx^N \\   \nonumber
&=& \frac{R^2}{z^2} e^{2 \tilde \vp(z)} \left( \eta_{\mu \nu} dx^\mu dx^\nu - dz^2\right).
\enqa
The effective action is
\begin{multline}
\label{action3}
\tilde S_{\it eff} = \int d^{d} x \,dz \,\sqrt{\vert \tilde g \vert} 
\, \tilde g^{N_1 N_1'} \cdots  \tilde g^{N_J N_J'}   \Big(  \tilde g^{M M'} D_M \Phi^*_{N_1 \dots N_J}\, D_{M'} \Phi_{N_1 ' \dots N_J'}  \\
 - \tilde \mu_{\it eff}^2(z)  \, \Phi^*_{N_1 \dots N_J} \, \Phi_{N_1 ' \dots N_J'} \Big),
 \end{multline}
where  $\sqrt{\vert \tilde g \vert} = (R\,  e^{\tilde \vp(z)}/z)^{d+1}$ and the effective mass $\tilde \mu_{\it eff}(z)$ is an {\it a priori} unknown function which encodes  kinematical aspects, but its $z$-dependence is needed  to avoid mixing between  kinematical and dynamical effects.

The use of warped metrics is useful to visualize the overall confinement behavior
by  following  an object in warped AdS space as it falls to the infrared region by the effects of gravity. The gravitational potential energy for an object of mass $M$ in general relativity is
given in terms of  the time-time component of the metric tensor $g_{00}$
\begin{equation} \label{V}
V = M c^2 \sqrt{\tilde g_{00}} = M c^2 R \, \frac{e^{ \tilde \vp(z)}}{z};
\end{equation}
thus, we may expect a potential that has a minimum at the hadronic scale $z_0 \sim 1/ \Lambda_{\rm QCD}$ and grows fast for larger values of $z$ to confine effectively a particle in a hadron within distances $z \sim z_0$. In fact, 
according to Sonnenscheim~\cite{Sonnenschein:2000qm}, a background
dual to a confining theory should satisfy the conditions for the
metric component $g_{00}$ 
\begin{equation}\label{Scond}
\partial_z (g_{00}) \vert_{z=z_0} = 0 , ~~~~ g_{00} \vert_{z = z_0} \ne 0,
\end{equation}
to display the Wilson loop area law for confinement of strings.

As in the case of the dilaton, considerable simplification  is brought by the introduction of  fields with tangent indices using a local Lorentz frame
\beq \label{inertialw}
  \hat  \Phi_{N_1 \dots N_J}=\left(\frac{z}{R}\right)^J\, e^{- 2 J \tilde \vp(z)}
 \,  \Phi_{N_1 \dots N_J}.
\enq
As shown in Appendix  \ref{wrapmet}, the action with a warped metric (\ref{action3}) and the effective action with a dilaton field (\ref{action2}) lead to identical results for the equations of motion for arbitrary spin, Eqs.  (\ref{PhiJ}) or (\ref{PhiJM}), provided that we identify the metric warp factor $\bar \vp(z)$ in (\ref{gw}) with the dilaton profile $\vp(z)$  according to $\tilde \vp(z) =   \vp(z)/(d-1)$, and 
\beq \label{mueffw}
\left(\tilde \mu_{\it eff}(z) R\right)^2 =  \left( m^2 + J z \frac{\tilde \vp '(z)}{d-1} - J z^2 \tilde \Omega^2(z) - J(d-J)  \right) e^{-2 \tilde \vp(z)},
\enq
where $\tilde \Om(z)$ is the warp factor of the affine connection for the metric (\ref{gw}), $\tilde \Om(z) = 1/z - \pa_z \tilde \vp$. 
A hadronic  spin-$J$ mode  propagating in the warped metric \req{gw} is normalized according to
 \begin{equation}  \label{Phinormw}
R^{d - 1 - 2 J} \int_0^{\infty} \! \frac{dz}{z^{d -1 - 2 J}} \, e^{(d - 1 - 2 J) \tilde \vp(z)} \Phi_J^2 (z) =
R^{d - 1} \int_0^{\infty} \! \frac{dz}{z^{d -1}} \, e^{(d - 1) \tilde \vp(z)} \hat \Phi_J^2 (z) = 1,
\end{equation}
in agreement with the normalization given in Ref.~\cite{Hong:2004sa}.

\section{Light-Front Holographic Mapping for Integer Spin\label{LFmapM}}

According to Dirac's classification of the forms of relativistic dynamics~\cite{Dirac:1949cp}, the fundamental generators of the Poincar\'e group
can be separated into kinematical and dynamical generators. In the light-front the kinematical generators  act along the initial surface and leave the light-front plane invariant: they are thus independent of dynamics and therefore contain no interactions. The dynamical generators change the light-front position and consequently depend
on  the interactions.

A physical hadron in four-dimensional Minkowski space has four-momentum $P_\mu$ and invariant
hadronic mass squared $P_\mu P^\mu = M^2$ which is determined by the
Lorentz-invariant Hamiltonian equation for the relativistic bound-state system
\begin{equation} \label{LFH}
H_{LF} \vert  \psi(P) \rangle =  M^2 \vert  \psi(P) \rangle,
\end{equation}
with  $H_{LF} = P_\mu P^\mu  =  P^- P^+ -  \mbf{P}_\perp^2$, and generators $P = (P^-, P^+,  \mbf{P}_\perp)$  constructed canonically from the QCD Lagrangian~\cite{Brodsky:1997de}.  The LF Hamiltonian $P^-$ generates LF time translations $i \hbar \frac{\pa}{\pa \tau} \vert \Psi \rangle = P^- \vert \Psi \rangle$
to evolve the initial conditions to all space-time,
whereas the LF longitudinal  $P^+$ and  transverse momentum $\mbf{P}_\perp$  are kinematical generators.
In addition to $P^+$ and $\mbf{P}_\perp$, the kinematical generators in the light-front frame are the $z$-component of the angular momentum $J^z$ and the boost operator $\mbf{K}$.  In addition to the Hamiltonian $P^-$,  $J^z$ and $J^y$ are also  dynamical generators. The light-front frame has the maximal number of kinematical generators~\cite{Dirac:1949cp}.

A remarkable correspondence between the equations of motion in AdS  and the Hamiltonian equation for relativistic bound-states (\ref{LFH}) was found in Ref.~\cite{deTeramond:2008ht}.   In fact, to a first semiclassical approximation, light-front QCD  is formally equivalent to the equations of motion on a fixed gravitational background~\cite{deTeramond:2008ht} asymptotic to AdS$_5$, where confinement properties  are encoded in the dilaton profile $\varphi(z)$ (\ref{action2}) which breaks the maximal symmetry of  AdS space.  For certain applications it is useful to reduce the multiparticle eigenvalue problem (\ref{LFH}) to a single equation~\cite{Pauli:1998tf,Brodsky:2012je}, instead of diagonalizing the Hamiltonian. The central problem then becomes the derivation of the effective interaction of the semiclassical
light-front Schr\"odinger equation  which acts only on the valence sector of the theory and has, by definition, the same eigenvalue spectrum as the initial Hamiltonian problem. For carrying out this program one must systematically express the higher Fock components as functionals of the lower ones. The method has the advantage that the Fock space is not truncated and the symmetries of the Lagrangian are preserved~\cite{Pauli:1998tf}.

In the limit of zero quark masses the longitudinal modes decouple from (\ref{LFH}) and
the LF eigenvalue equation $P_\mu P^\mu \vert \phi \rangle  =  M^2 \vert \phi \rangle$
is thus a light-front  wave equation for $\phi$~\cite{deTeramond:2008ht}
\begin{equation} \label{LFWE}
\left(-\frac{d^2}{d\zeta^2}
- \frac{1 - 4L^2}{4\zeta^2} + U\left(\zeta, J\right) \right)
\phi_{J,L,n}(\zeta^2) = M^2 \phi_{J,L,n}(\zeta),
\end{equation}
a relativistic single-variable  LF  Schr\"odinger equation~\cite{Brodsky:2012je}.  The boost-invariant transverse-impact variable $\zeta$~\cite{Brodsky:2006uqa}
measures the separation of quark and gluons at equal light-front time, and  it also allows one to separate the bound-state dynamics
of the constituents from the kinematics of their internal angular momentum~\cite{deTeramond:2008ht}. For a two-parton bound state
\beq \label{zeta2}
\zeta = \sqrt{x(1-x)} \vert \mbf{b}_\perp \vert,
\enq
where $x$ is the longitudinal momentum fraction and $ \mbf{b}_\perp$ is  the transverse-impact distance between the two quarks.
In first approximation, the effective interaction $U$ is instantaneous in LF time and acts on the lowest state of the LF Hamiltonian.  This equation describes the spectrum of mesons as a function of $n$, the number of nodes in $\zeta$, the total angular momentum  $J$, which represent the maximum value of $\vert J^z \vert$, $J = \max \vert J^z \vert$,
and the internal orbital angular momentum of the constituents $L= \max \vert L^z\vert$.

Factoring out the scale factor $(1/z)^{ J - (d-1)/2}$ and the dilaton factor from the AdS field we write
\begin{equation} \label{Phiphi}
\Phi_J(z)   = \left(\frac{R}{z} \right)^{ J - (d-1)/2 } e^{- \varphi(z)/2} \, \phi_J(z) .
\end{equation}
Upon the substitution of the holographic variable $z$ by the light-front invariant variable $\zeta$ and replacing \req{Phiphi}
into the AdS wave eigenvalue equation  (\ref{PhiJM}), we find for $d=4$ the QCD light-front frame-independent wave equation (\ref{LFWE}) with effective potential~\cite{deTeramond:2010ge}
\begin{equation} \label{U}
U(\zeta, J) = \half \varphi''(\zeta) +\frac{1}{4} \varphi'(\zeta)^2  + \frac{2J - 3}{2 \zeta} \varphi'(\zeta) ,
\end{equation}
provided that the fifth dimensional AdS mass $m$ in (\ref{PhiJM}) is related to the light-front internal orbital angular momentum $L$  and the total angular momentum $J$ of the hadron according to
\begin{equation} \label{muRJL}
(m R)^2 = - (2-J)^2 + L^2.
\end{equation}
Light-front holographic mapping thus implies that the fifth  AdS mass  $m$ in (\ref{PhiJM})  is not a free parameter  but scales according to (\ref{muRJL}), thus giving a 
precise expression for the AdS effective mass $\mu_{\it eff}(z)$ in (\ref{action2}).
The light-front mapping  provides the basis for a profound connection between physical QCD formulated in the light-front  and the physics of hadronic modes in AdS space. However, important differences are also apparent:  Eq. (\ref{LFH}) is a linear quantum-mechanical equation of states in Hilbert space, whereas Eq. (\ref{PhiJM}) is a classical gravity equation; its solutions describe spin-$J$ modes propagating in a higher dimensional
warped space. Physical hadrons are composite, and thus inexorably endowed of orbital angular momentum. Thus, the identification
of orbital angular momentum is of primary interest in establishing a connection between the two approaches.

If $L^2 < 0$, the LF Hamiltonian   is unbounded from below
$\langle \phi \vert P_\mu P^\mu \vert \phi \rangle <0$  and the spectrum contains an
infinite number of unphysical negative values of $M^2 $ which can be arbitrarily large.
As $M^2$ increases in absolute value, the particle becomes
localized within a very small region near $\zeta = 0$, since  the effective potential is conformal at small  $\zeta$.
For $M^2 \to - \infty$ the particle is localized at $\zeta = 0$, the
particle ``falls towards the center''~\cite{LL:1958}.
The critical value  $L=0$  corresponds to the lowest possible stable solution, the ground state of the light-front Hamiltonian.
For $J = 0$ the five dimensional mass $m$
 is related to the orbital  momentum of the hadronic bound state by
 $(m R)^2 = - 4 + L^2$ and thus  $(m R)^2 \ge - 4$. The quantum mechanical stability condition $L^2 \ge 0$ is thus equivalent to the Breitenlohner-Freedman stability bound in AdS~\cite{Breitenlohner:1982jf}.
The scaling dimensions are $2 + L$, independent of $J$, in agreement with the
twist-scaling dimension of a two-parton bound state in QCD~\cite{Brodsky:1973kr}.
It is important to notice that in the light-front the $SO(2)$ Casimir for orbital angular momentum $L^2$
is a kinematical quantity,  thus giving a kinematical interpretation of the AdS mass.
In contrast, the usual $SO(3)$ Casimir $L(L+1)$ from non-relativistic physics  is
rotational, but not boost invariant.

\subsection{A Hard- and Soft-Wall Model for Mesons \label{softwallmesons}}

The simplest holographic example is a truncated model where quarks propagate freely in the
hadronic interior up to the confinement scale, whereas the confinement dynamics is included by the boundary conditions at  $1/\Lambda_{\rm QCD}$~\cite{Polchinski:2001tt}.
This model provides an  analog of the MIT bag model~\cite{Chodos:1974je}
where quarks are permanently confined inside a finite region of space. 
In contrast to bag models, boundary conditions are imposed on the 
boost-invariant variable $\zeta$, not on the bag radius at fixed time.  
The resulting model is a manifestly Lorentz invariant model
with confinement at large distances, while incorporating conformal
 behavior at small physical separation.  The eigenvalues of the  LF wave equation (\ref{LFWE}) for the hard-wall model ($U = 0$) are determined by the boundary conditions $\phi(z = 1/\Lambda_{\rm QCD}) = 0$, and are given in terms of the roots  $\beta_{L,k}$ of the Bessel functions: $\mathcal{M}_{L,k} = \beta_{L,k} \Lambda_{\rm QCD}$. By construction, the hard wall model  has  a simple separation of kinematical and dynamical aspects, but it has shortcomings 
 when trying to describe the observed meson spectrum~\cite{deTeramond:2012rt}.  The model 
 fails to account for the pion as a chiral $M=0$ state and it is degenerate with respect to the orbital quantum number $L$, thus leading to identical trajectories for pseudoscalar and vector mesons. It also fails to account for the important   splitting for the $L=1$ $a$-meson states for different values of $J$. Furthermore,  for higher quantum excitations the spectrum behaves as $M \sim 2n + L$, in contrast to the usual Regge dependence $M^2 \sim n + L$ found experimentally~\cite{Klempt:2007cp}. As a consequence, the radial modes are not well described in the truncated-space model.

The shortcomings of the hard-wall model are evaded with the soft wall model~\cite{Karch:2006pv} where the sharp cutoff is modified by a 
a dilaton profile $\vp(z) = \lambda z^2$. The soft-wall model 
leads to linear Regge trajectories~\cite{Karch:2006pv} and avoids the ambiguities in the choice of boundary conditions at the infrared wall.  
In fact, it can be shown that if one starts with a dilaton of the general form $\varphi(z, s) = \lambda z^s$, for arbitrary values of  $s$, the constraints imposed by chiral symmetry in the limit of massless quarks determine uniquely the value  $s = 2$~\cite{Brodsky:2013yy}. This is a remarkable result, since this value corresponds precisely to the dilaton profile required to  reproduce the  linear Regge behavior.

From (\ref{U}) we obtain  the effective potential
\beq \label{Ulambda}
U(\zeta) =   \la^2 \zeta^2 + 2 \la (J - 1),
\enq
which  corresponds  to a transverse oscillator in the light-front.
For the effective potential (\ref{Ulambda}) equation  (\ref{LFWE}) has eigenfunctions
\beq \label{phi}
\phi_{n, L}(\zeta) = \la^{(1+L)/2} \sqrt{\frac{2 n!}{(n\!+\!L\!)!}} \, \zeta^{1/2+L}
e^{- \vert \la \vert \zeta^2/2} L^L_n(\vert \la \vert \zeta^2) ,
\enq
and eigenvalues
\beq
M^2 = \left( 4n + 2 L + 2\right) \vert \la \vert + 2 \la(J-1).
\enq
The LF wavefunctions $\phi(\zeta) = \langle \zeta \vert \phi \rangle$ are normalized as $\langle \phi \vert \phi \rangle = \int d \zeta \, \phi^2(z) = 1$ in accordance with (\ref{Phinorm}).

Except for $J=1$ the spectrum predictions are significantly different for $\la > 0$ or  $\la < 0$.   The predicted spectrum for  $\la>0$
\beq\label{M2SFM}
\mathcal{M}_{n, J, L}^2 = 4 \la \left(n + \frac{J+L}{2} \right),
\enq
gives a very good description of the excitation spectrum of the mesons~\cite{deTeramond:2012rt}. In particular, the lowest possible solution for $n = L = J =0$ has eigenvalue $M^2 =0$.
This is a chiral symmetric bound state of two massless quarks and scaling dimension 2, which we identify with the lowest state, the pion.  Furthermore, the model with $\la>0$
accounts for the mass pattern observed in radial and orbital excitations, as well as for the triplet splitting for the $L=1$, $J = 0, 1, 2$, vector meson $a$-states~\cite{deTeramond:2012rt}.  The slope of the Regge trajectories gives a value $\la \simeq 0.5 ~{\rm GeV}^2$.  
The  result \req{M2SFM} was found in Ref. \cite{Gutsche:2011vb}.

On the other hand, the solution for $\la<0$  leads to a pion mass heavier than the $\rho$ meson and a meson spectrum given by $M^2 = 4 \lambda \, (n + 1 + (L - J)/2$, in clear disagreement with the observed spectrum. 
Thus  the solution $\la<0$ is incompatible with the light-front  constituent interpretation of hadronic states.  Since the confining term $\la^2 \zeta^2$ in the effective
potential \req{Ulambda} does not depend on the sign of $\lambda$ it is
always possible to compensate a change of the sign of $\lambda$
without changing the spectrum by adding {\it ad hoc}
$z$-dependent mass terms to the Lagrangian~\cite{Gutsche:2011vb}.
We note that in our approach, however, the $z$-dependent mass
terms are uniquely fixed.  Other possible approaches are discussed in \cite{HGD12}, but those are shown to give a worse description of the data.

The solution $\la>0$ is consistent with the Wilson loop area law condition (\ref{Scond}) with a minimum $z_0  \sim 1/ \sqrt{\la}$. In fact,  the corresponding modified metric  for the soft-wall model can be interpreted in the higher dimensional warped AdS space as a gravitational potential in the fifth dimension \req{V}
\begin{equation} 
V(z) =  M c^2 R \, \frac{e^{\lambda z^2/3}}{z}  .
\end{equation}
For $\la < 0$
the potential decreases monotonically, and thus an object located in the boundary of AdS space will fall to infinitely large 
values of $z$.  This is illustrated in detail by Klebanov and Maldacena in Ref.  \cite{Klebanov:2009zz}.
For $\la > 0$, the potential is nonmonotonic and has an absolute minimum at $z_0 \sim 1/\sqrt{\la}$.  
Furthermore, for large values of $z$ the gravitational potential increases exponentially, thus confining any object  to distances 
$\langle z \rangle \sim 1/ \sqrt{\la}$~\cite {Andreev:2006vy, deTeramond:2009xk}.

In the model discussed in Ref.~\cite{Karch:2006pv} higher spin equations are constructed by imposing invariance of the AdS action under gauge transformations. This implies setting the fifth dimensional mass equal to zero. This construction needs a negative value for $\la$ and is  incompatible with the light-front constituent interpretation of the gauge/gravity duality, since the light front-mapping implies the kinematical constraint (\ref{muRJL}), thus fixing $L$ for a given $J$. For example, for the $\rho$ meson $J=1$, and the only allowed value would be $L=1$. This would exclude its main $L=0$ component.

Finally, we notice that for $m = M=0$ the AdS wave equation for bound states \req{PhiJM} reduces to
 \beq 
\partial_z \left(\frac{e^{\varphi(z)}}{z^{d-1- 2J}} \partial_z\right) \Phi_J(z) =0,
\enq 
and has the solution
$\Phi_J = C \int_a^z dz \, e^{-\vp(z')}  z^{ d-1-2J}$.
For $d=4$, $J=1$ and the dilaton profile $\vp(z)= \la \, z^2$ this leads to the regular solution 
\beq
 \label{bads} \Ph_{J=1}(z) = A\, e^{-\la \,z^2} +B,
  \enq
with arbitrary constants $A$ and $B$. The existence of such a solution, which for $B=0$ decays exponentially, has been used in Ref.  \cite{Karch:2010eg} as an argument against a positive dilaton profile $\la>0$, since it would correspond to a normalizable wave function for a massless vector meson.  However, if one uses the correct
measure \req{Phinorm} it becomes clear that the  normalization integral  \req{Phinorm} with the solution \req{bads} diverges either at $z =0$ or $z \to \infty$; thus \req{bads} does not represent a physical bound state in light-front holographic QCD.

\section{Half-Integer Spin \label{half}}

The study of the  internal structure and excitation spectrum of baryons is one of the most challenging aspects of hadronic physics.
An important goal of computations in lattice QCD is the reliable extraction of the excited nucleon mass spectrum.
Lattice calculations of the ground state light hadron masses agree well with experimental values~\cite{Dudek:2011zz}. However, the
excitation spectrum of the nucleon represents a formidable challenge to lattice QCD due to the enormous computational complexity required for the extraction of meaningful data beyond the leading ground state configuration~\cite{Edwards:2011jj}.  Moreover, a large basis of interpolating operators is required  since excited nucleon states are classified according to  irreducible representations of the lattice, not the total angular momentum. In contrast, the semiclassical light-front holographic wave equation (\ref{LFWE})  describes Lorentz frame-independent relativistic bound states at equal light-front time  with an analytic simplicity comparable to the Schr\"odinger equation of atomic physics at equal instant time.
It is therefore tempting to extend basic gauge/gravity ideas to describe excited baryons as well, by considering the propagation of higher-spin Dirac modes in AdS space and the mapping of the corresponding wave equations to the light front in physical-space time.

In the usual AdS/CFT correspondence the baryon is an $SU(N_C)$ singlet bound state of $N_C$ quarks in the large $N_C$ limit. Since  there are no quarks in this theory, quarks are introduced as external sources at the AdS asymptotic boundary~\cite{Witten:1998xy, Gross:1998gk}.  The baryon is constructed  as an $N_C$ baryon vertex located in the interior of AdS. In this top-down string approach baryons are usually described as solitons
 or Skyrmion-like objects~\cite{Hong:2007kx, Hata:2007mb}.   In contrast, the light-front holographic approach  is based on the precise mapping of AdS expressions to light-front QCD.  Consequently, we will construct baryons corresponding to $N_C=3$ not $N_C \to  \infty$.
 We would expect that in the limit of zero quark masses we  find a relativistic bound state  light-front wave equation with a geometrical equivalent to the equation of  motion for a higher half-integral hadronic state in a warped AdS space-time.  As it turns out, the analytical exploration of the baryon spectrum using gauge/gravity duality ideas is not as simple, or as well understood, as the meson case, and further work beyond the scope of the present article is required.  However, as we shall discuss below,   even a relatively simple approach
provides a framework for a useful analytical exploration of the strongly coupled dynamics of baryons  which gives important insights into the systematics of the light baryon spectrum using simple analytical methods.

\subsection{Invariant Action and Equations of Motion}

Fields with half-integer spin $J = T + \half$ are conveniently described by Rarita-Schwinger spinors~\cite{Rarita:1941mf},  $\left[ \Psi_{N_1 \cdots N_T}\right]_\al$,  objects which transform as  
symmetric tensors of rank $T$ with indices $N_1 \dots N_T$, and as  Dirac spinors with index $\alpha$.
The Lagrangian of fields with arbitrary half-integer spin in a higher-dimensional space is vastly complex.  General covariance allows for a superposition of terms
 of the form
$$\bar \Psi_{N_1 \dots N_T} \Ga^{[N_1 \dots N_T M N_1' \dots N_T']} D_M  \Psi_{N'_1\ \dots N_T'}, $$
and mass terms
$$\mu \bar \Psi_{N_1 \dots N_T} \Ga^{[N_1 \dots N_T N_1' \dots N_T']}   \Psi_{N_1 \dots N_T'}, $$
where the tensors $\Ga^{[\cdots]}$ are antisymmetric products of Dirac  matrices and a sum over spinor indices is  understood. The maximum number of independent Dirac matrices depends on the dimensionality of space. In Appendix \ref{threehalf} we present explicitly the case of spin $\frac{3}{2}$.

In flat space, the equations describing a free particle with spin $T+\frac{1}{2}$ are~\cite{Rarita:1941mf} 
\beq \label{RS}
 \left(i \ga^\mu \pa_\mu - M\right) \Psi_{\nu_1 \cdots \nu_T} =0,  \qquad {\ga^\nu} \Psi_{\nu \nu_2 \cdots \nu_T}=0.
  \enq
 The subsidiary conditions of the integral spin theory for the $T$ tensor indices  \req{scPhi}
 \beq \label{scPsi}
 \eta^{\mu \nu} \pa_\mu  \Psi_{\nu \nu_2 \cdots \nu_T}=0, \quad
\eta^{\mu \nu}   \Psi_{ \mu \nu \nu_3  \cdots \nu_T}=0, 
\enq 
are a consequence of these equations~\cite{Rarita:1941mf}.

We have seen in Sec.~\ref{integer} that the kinematical subsidiary conditions for fields with integer spin in $d$-dimensional space  follow from the simple effective action \req{action2}. The actual form of the Dirac equation for Rarita-Schwinger spinors  (\ref{RS}) in flat space-time motivates us to start with a simple  effective action for arbitrary half-integer spin in AdS space which, in the absence of dynamical terms, preserves maximal symmetry of AdS  in order to describe the correct kinematics.   We  also expect that the effective action for higher half-integer spins in AdS space will  also lead to the  Rarita-Schwinger condition 
${\ga^\nu} \Psi_{\nu \nu_2 \cdots \nu_T}=0$  in physical space-time.

We will start with an effective action in AdS$_{d+1}$ motivated by \req{RS} including a dilaton term $\vp(z)$ and an effective interaction $\rho(z)$ (See also Ref. \cite{Gutsche:2011vb})
 \begin{multline} \label{af} 
~~~  S_{F \it  eff} = \half  \int  \!
d^{d} x \,dz\,  \sqrt{| g|} \, e^{\vp(z)}  g^{N_1\,N_1'} \cdots g^{N_T\,N_T'}  \\   
\left[ \bar  \Psi_{N_1 \cdots N_T}  \Big( i \, \Ga^A\, e^M_A\,  D_M -  \mu - \rho(z)\Big)
 \Psi_{N_1' \cdots N_T'} + h.c. \right] , ~~~~~~
 \end{multline}
where  $\sqrt{g} = \left(\frac{R}{z}\right)^{d+1}$ and $e^M_A$ is the inverse vielbein, $e^M_A = \left(\frac{z}{R}\right) \delta^M_A$. The covariant derivative $D_M$ of a Rarita-Schwinger spinor  includes the affine connection and the spin connection (Appendix \ref{intspinB}),
 and the tangent-space Dirac matrices obey the usual anti-commutation relation $\left\{\Gamma^A, \Gamma^B\right\} = \eta^{A B}$. For 
$\vp(z) = \rho(z) = 0$ the effective action  (\ref{af}) preserves the maximal symmetry of AdS space.
The reason why we need to introduce  an additional  symmetry breaking term $\rho(z)$  in \req{af} will become clear soon.
As we shall show below, this action indeed contains the Rarita-Schwinger condition given in (\ref{RS}) and the subsidiary conditions \req{scPsi}.

We will confine ourselves to the physical polarizations orthogonal to the holographic dimension
\beq \Psi_{z N_2 \dots N_T}=0 \label{orthf}, \enq
and obtain the equations of motion from the Euler-Lagrange 
equations in the subspace defined by (\ref{orthf})
 \beq \label{ELJf} 
 \frac{ \de S_{F  \it eff}}{ \de \bar \Psi_{\nu_1 \nu_2 \cdots \nu_J}} = 0,
\enq
 and
 \beq \label{ELzf}
  \frac{\de S_{F \it eff}}{\de \bar \Psi_{ z N_2 \cdots N_J}} = 0.
 \enq

Our derivation of the half-integer spin theory follows the lines along Sec. \ref{integer}. 
We introduce fields with tangent indices using a local Lorentz frame as in  \req{inertial}
 \beq 
 \hat \Psi_{N_1 \dots N_T}  = \left(\frac{z}{R}  \right)^T \Psi_{N_1\dots N_T},
 \enq
 and use the results of Appendix \ref{intspinB} to separate
 the  action into a part $S_{F \it eff}^{[0]}$ containing only spinors
orthogonal to the holographic direction, and a term $S_{F \it eff}^{[1]}$, containing  terms linear in
$\bar \Psi_{z N_2 \dots N_T}$;  the remainder does not contribute to
the Euler-Lagrange equations \req{ELzf}. Since the fermion action is
linear in the derivatives,  the calculations are considerably
simpler compared with the integer spin case,  and one obtains
 \begin{multline}\label{afnu}
 S_{F \, \it eff}^{[0]} =   \int d^{d} x \,dz\, \Big(\frac{R}{z}\Big)^{d+1} e^{\vp(z)}  \eta^{\nu_1 \nu_1'}\dots \eta^{\nu_T \nu_T'}
 \bigg( \frac{i}{2} \,e^M_A \, \,\overline  {\hat \Psi}_{\nu_1 \cdots \nu_T} \Ga^A \,\pa_M
{ \hat \Psi}_{\nu'_1 \dots \nu'_T}   \\
   - \frac{i}{2}\, e^M_A  \left(\pa_M \overline {\hat \Psi}_{\nu_1 \dots \nu_T}\right)  \Ga^A\, { \hat \Psi}_{\nu'_1 \cdots \nu'_T } 
- \left( \mu + \rho(z) \right)   \overline {\hat \Psi}_{\nu_1 \cdots \nu_T} {\hat \Psi}_{\nu_1' \cdots \nu_T' }
 \bigg),
\end{multline}
and
  \begin{multline} \label{afz}
 S_{F \, \it eff}^{[1]} = -  \int d^d x \,dz\, \Big(\frac{R}{z}\Big)^{d}  e^{ \vp(z)}   \eta^{N_2 N_2'}\cdots  \eta^{N_T  N_T'} \\
 T \,\Om(z) \,
  \Big( \overline{ \hat \Psi}_{z N_2 \cdots N_T}  \,\Gamma^\mu \hat \Psi_{\mu
  N_2' \cdots N_T'}
 + \overline{ \hat \Psi}_{\mu  N_2 \cdots N_T} \,  \Gamma^\mu \hat \Psi_{z
  N'_2 \dots N'_T}\Big),
 \end{multline}
where the factor of the affine connection, see  Eqs. \req{gammads}   and \req{afc}, is $\Om(z) = 1/z$.

Performing a partial integration, the  action \req{afnu} becomes:
  \begin{multline} \label{afp}
   S_{F \,\it eff}^{[0]}  =  \int d^d x \, dz \, \Big(\frac{R}{z}\Big)^{d} \,e^{\vp(z)}   \,
\eta^{\nu_1 \nu_1'}\cdots \eta^{\nu_T \nu_T'}  \\
 \overline{\hat \Psi}_{\nu_1 \cdots \nu_T}\, \Big( i \et^{NM}  \Ga_M \pa_N +
 \frac{i}{2 z}\Ga_z  \left(d -  z \vp'(z) \right) -  \mu R- \rho(z)   \Big) \hat \Psi_{\nu'_1 \dots \nu'_T} ,
\end{multline}
plus surface terms.

The variation of  \req{afz}  yields indeed
  the Rarita-Schwinger condition in physical space-time \req{RS} 
 \beq \label{dirac-SE1}
 \gamma^\nu \hat \Psi_{\nu  \nu_2 \, \dots \,\nu_T} =0, \enq
and the variation of  \req{afp} provides the AdS Dirac-like wave equation
\beq \label{DEhat}
\left[ i \left( z \eta^{M N} \Gamma_M \partial_N + \frac{d - z \vp'}{2} \Gamma_z \right) - \mu R - R \, \rho(z)\right]   \hat \Psi_{\nu_1 \dots \nu_T}=0.
\enq

Although the dilaton term $\vp'(z)$ shows up in the equation of motion  \req{DEhat}, it actually does not lead to dynamical effects, since it  can be absorbed by rescaling  the Rarita-Schwinger spinor according to $ {\tilde \Psi}_{\nu_1 \dots \nu_T} = e^{\vp(z)/2} \hat\Psi_{\nu_1 \dots \nu_T} $. 
This leads to the equation
\beq \label{Psihat}
\left[ i \left( z \eta^{M N} \Gamma_M \partial_N + \frac{d}{2} \Gamma_z \right) - \mu R - R \, \rho(z)\right]   \tilde \Psi_{\nu_1 \dots \nu_T}=0.
\enq
Thus, for fermion fields in AdS one cannot introduce confinement by the introduction of a dilaton in the action since it can be rotated away~\cite{Kirsch:2006he}.
This is a consequence of the linear covariant derivatives in the fermion action, which also prevents a mixing between dynamical and kinematical effects, and thus, in contrast with the effective action for integer spin fields \req{action2}, the AdS mass $\mu$  in Eq. \req{af} is constant.
As a result, one must introduce an effective confining interaction $\rho(z)$ in the fermion action to break conformal symmetry and generate a baryon spectrum~\cite{Brodsky:2008pg, Abidin:2009hr}.  
This interaction can be constrained by the condition that the `square' of the Dirac equation leads to a potential which matches the dilaton-induced potential for integer spin.

Going back from the tangential space coordinates to covariant tensors and scaling away the dilaton factor in  \req{af} by a field redefinition
\beq
\Psi \to e^{\vp(z)/2} \Psi
\enq
 we obtain
\beq \label{PsiT}
\left[ i \left( z \eta^{M N} \Gamma_M \partial_N + \frac{d - 2 T}{2} \Gamma_z \right) - \mu R - R \, \rho(z)\right]   \Psi_{\nu_1 \dots \nu_T}=0,
\enq
which is the half-integral spin equivalent of Eqs. (\ref{PhiJ}), 
and the Rarita-Schwinger condition
\beq \label{dirac-SE}
 \gamma^\nu \Psi_{\nu  \nu_2 \, \dots \,\nu_T} =0. \enq
In fact,  the Rarita-Schwinger condition in the physical subspace of AdS spinors \req{dirac-SE} in flat four-dimensional space also 
entails, with the extended Dirac equation \req{DEhat},  the subsidiary conditions for the tensor indices required to eliminate the lower spins.  Thus multiplying Eq. \req{DEhat}  by $\ga^{\nu}$ and using  \req{dirac-SE} we obtain
\beq
i\,z \,\eta^{MN}  \,\ga^{\nu}  \,\Ga_M \pa_N \, \Psi_{\nu \nu_2 \cdots \nu_T} = 0,
\enq
and
\beq
i\,z \,\eta^{MN}  \, \Ga_M  \,\ga^{\nu}  \,\pa_N \, \Psi_{\nu \nu_2 \cdots \nu_T} = 0 .
\enq
Adding the last  two equations  and making use of the symmetry of the tensor indices of the Rarita-Schwinger spinors we get the condition
\beq
2 i\,z \,\eta^{\nu  N}  \,\pa_N \, \Psi_{\nu  \cdots \nu_T} =0,
\enq
which gives indeed the divergence condition is Eqs. \req{scPsi}: $ \eta^{\mu \nu} \pa_\mu  \Psi_{\nu \nu_2 \cdots \nu_T}=0$.
The derivation of the trace condition  is exactly the same as in flat space.
From \req{dirac-SE} it follows that
$
\ga^{\nu} \, \ga^{\mu} \Psi_{\mu \nu \nu_3 \cdots \nu_T}=0,
$ from which  the trace condition in  \req{scPsi}  is obtained from the symmetry of the indices of the spinor field: 
$
\eta^{\mu \nu} \Psi_{\mu \nu \nu_2 \cdots \nu_T}=0$.
 We compare our results from the effective action \req{af} for spin-$\frac{3}{2}$  with the results from Refs. \cite{Volovich:1998tj} and \cite{Matlock:1999fy} in Appendix \ref{threehalf}.
 
 Identical results for the equations of motion for arbitrary half-integer spin are obtained if one starts with the distorted metric \req{gw}.  One finds that the effective fermion action with a dilaton field \req{af} is equivalent to the fermion action with warped metrics, provided that we identify the dilaton profile according to $\tilde \vp(z) = \vp(z)/d$ and the effective mass 
 $\tilde \mu(z)$ in the warped action with the mass $\mu$ in \req{af} according to $\tilde \mu(z) = e^{- \tilde \vp(z)} \mu$. Thus, one cannot introduce confinement in the effective AdS action  for fermions either by a dilaton profile or by additional warping of the AdS metrics in the infrared. In each case one requires  an additional effective interaction as introduced in the effective action \req{af} with $\rho(z) \ne 0$.

\section{Light-Front Holographic Mapping for Half-Integer Spin\label{LFmapB}}

One can also take as starting 
point the construction of light-front wave equations in physical space-time for baryons by studying  the LF transformation properties of spin-$\half$ states~\cite{{Brodsky:2008pg}}. The light-front wave equation describing baryons is a matrix eigenvalue equation $D_{LF} \vert \psi \rangle = \mathcal{M} \vert \psi \rangle$ with $H_{LF} = D_{LF}^2$. In a $2 \times 2$ chiral spinor component representation,  the light-front equations are given by the coupled linear differential equations
\begin{eqnarray} \label{LFDE}  \nonumber
- \frac{d}{d\zeta} \psi_-  - \frac{\nu+\half}{\zeta}\psi_-  -  V(\zeta) \psi_-&=& M \psi_+ , \\
 \frac{d}{d\zeta} \psi_+ - \frac{\nu+\half}{\zeta}\psi_+  - V(\zeta) \psi_+ &=& M \psi_- , 
\end{eqnarray}
where the invariant variable $\zeta$ for an $n$-parton bound state  is the $x$-weighted 
transverse impact variable of the $n-1$ spectator system~\cite{Brodsky:2006uqa},
\beq  \label{hologz}
\zeta = \sqrt{\frac{x}{1-x}} \Big\vert \sum_{j=1}^{n-1} x_j \mbf{b}_{\perp j} \Big\vert,
\enq
and  $x = x_n$ is the longitudinal light-front momentum fraction of the active quark (For $n=2$ we recover \req{zeta2}).
As discussed below, we can identify $\nu$ with the light-front orbital angular momentum $L$,  $\nu = L+1$, the relative  angular momentum  between the active and the spectator cluster.

A physical baryon has plane-wave solutions  with four-momentum $P_\mu$,  invariant mass $P_\mu P^\mu = M^2$, and polarization indices along the physical coordinates. It thus  satisfies the Rarita-Schwinger equation for spinors in physical space-time \req{RS}
 \beq
  \left(i \ga^\mu \pa_\mu - M \right) u_{\nu_1 \cdots \nu_T}({P}) =0,  \qquad {\ga^\nu} u_{\nu \nu_2 \cdots \nu_T}({P})=0.
  \enq
 Factoring out  from the AdS spinor field $\Psi$ the four-dimensional plane-wave and spinor dependence, as well as the scale factor $(1/z)^{T-d/2}$, we write
\beq \label{Psipsi}
\Psi^\pm_{\nu_1 \cdots \nu_T}(z)   = e^{ i P \cdot x}   \left(\frac{R}{z} \right)^{T-d/2}   \psi^\pm_T(z) \, u^\pm_{\nu_1 \cdots \nu_T} ({P}),
\enq
where $T = J - \half$ and the chiral spinor  $u^\pm_{\nu_1 \dots \nu_T} =
\half (1 \pm \gamma_5)u_{\nu_1 \dots \nu_T}$  satisfies the four-dimensional chirality equations
 \beq
 \gamma \cdot P \, u^\pm_{\nu_1 \dots \nu_T}({P}) = M  u^\mp_{\nu_1 \dots \nu_T}({P}), \qquad
\gamma_5 u^\pm_{\nu_1 \dots \nu_T}({P}) = \pm \, u^\pm_{\nu_1 \dots \nu_T}({P}).
\enq

Upon replacing  the holographic variable $z$ by the light-front invariant variable $\zeta$ and substituting \req{Psipsi} into  the AdS wave equation \req{PsiT}  we recover its LF expression (\ref{LFDE}), provided that $ \vert \mu R \vert = \nu + \half $ and
$\psi^\pm_T = \psi_\pm$, independent of the value of $T = J - \half$.  We also find that the effective LF potential in the light-front Dirac equation \req{LFDE} is determined by the effective interaction $\rho(z)$ in the effective action \req{af},
\begin{equation}
V(\zeta) = \frac{R}{\zeta} \rho(\zeta),
\end{equation}
 which is a $J$-independent potential. This is a remarkable result, since it implies that independently of the specific form of the potential, the value of the baryon masses along a given Regge trajectory depends only on the LF orbital angular momentum $L$, and thus, in contrast with the vector mesons, there is no spin-orbit coupling, in agreement with the observed near degeneracy in the baryon spectrum~\cite{Klempt:2007cp}.   Equation (\ref{LFDE}) 
 is equivalent to the system of second order equations
 \begin{equation} \label{LFWEA}
\left(-\frac{d^2}{d\zeta^2}
- \frac{1 - 4 \nu^2}{4\zeta^2} + U^+(\zeta) \right) \psi_+ = \mathcal{M}^2 \psi_+,
\end{equation}
and
\begin{equation} \label{LFWEB}
\left(-\frac{d^2}{d\zeta^2}
- \frac{1 - 4(\nu + 1)^2}{4\zeta^2} + U^-(\zeta) \right) \psi_- = \mathcal{M}^2 \psi_-,
\end{equation}
where
\beq \label{Upm}
U^\pm(\zeta) = V^2(\zeta) \pm V'(z) + \frac{1 + 2 \nu}{\zeta} V(\zeta),
\enq
with $\nu = L + 1$.

For baryons, the corresponding
 interpolating operator for an $N_C = 3$ physical baryon
$ \mathcal{O}_{3 + L} =  \psi D_{\{\ell_1} \dots
 D_{\ell_q } \psi D_{\ell_{q+1}} \dots
 D_{\ell_m\}} \psi$,  $L = \sum_{i=1}^m \ell_i$, is a 
 twist-3,  dimension $9/2 + L$ with scaling behavior given by its
 twist-dimension $3 + L$.  We thus require $\nu  = L+1$ in order to match the short-distance scaling behavior.    
 Note that $L$ is the maximal value of $|L^z|$ in a given LF Fock state.  
An important feature of  bound-state relativistic theories  is that hadron eigenstates have in general Fock components with different $L$ components.  By convention one labels the eigenstate with its minimum value of $L$.  For example, the symbol $L$ in the light-front AdS/QCD spectral  prediction for mesons (\ref{M2SFM}) refers to the {\it minimum } $L$  (which also corresponds to the leading twist) and $J$ is the total angular momentum of the hadron.

\subsection{A Hard- and Soft-Wall Model for Baryons \label{softwallbaryons}}

As for the case of mesons,  the simplest holographic model of baryons is the hard-wall model, where confinement dynamics is included by the boundary conditions  at $z \simeq 1/\La_{\rm QCD}$.  To determine the boundary conditions we integrate by parts (\ref{af}) for $\varphi(z) = \rho(z) = 0$  and use the equations of motion.  We then find
\begin{equation}
S_F= - \lim_{\epsilon \to 0} \, R^d \int \frac{d^dx}{2 z^d} \Big ( \bar \Psi_+ \Psi_- - \bar \Psi_- \Psi_+ \Big) \Big \vert_\epsilon^{z_0},
\end{equation}
where  $\Psi_\pm = \half \left(1 \pm \gamma_5 \right) \Psi$.  Thus in a truncated-space holographic model, the light-front modes $\Psi_+$ or $\Psi_-$ should vanish at the boundary $z = 0$ and $z_0 = 1/\La_{\rm QCD}$. This condition fixes the boundary conditions and determines the baryon spectrum in the truncated hard-wall model~\cite{deTeramond:2005su}:
$M^+ = \beta_{\nu,k} \, \Lambda_{\rm QCD}$, and $M^- = \beta_{\nu+1,k} \, \Lambda_{\rm QCD}$,
 with a scale-independent  mass ratio determined by the zeros of Bessel functions $\beta_{\nu, k}$. Equivalent results follow from the hermiticity of the LF Dirac operator $D_{LF}$ in the eigenvalue equation  $D_{LF} \vert \psi \rangle = \mathcal{M} \vert \psi \rangle$. The orbital excitations of baryons in this model are approximately aligned along  two trajectories corresponding to even and odd parity states~\cite{deTeramond:2005su, deTeramond:2012rt}. The spectrum shows a clustering of states with the same orbital $L$, consistent with a strongly suppressed spin-orbit force.  As for the case for mesons, the hard-wall model predicts $\mathcal{M} \sim 2n + L$, in contrast to the usual Regge behavior $\mathcal{M}^2 \sim n + L$ found in experiment~\cite{Klempt:2007cp}.  The radial modes are also not well described in the truncated-space model. 
 
Let us now examine a model similar to the soft-wall dilaton model  for mesons by introducing an effective potential, which also leads to linear Regge trajectories in both the orbital and radial quantum numbers for baryon excited states.  As we have discussed,
a dilaton factor in the  fermion action can be scaled away by a field redefinition.   We thus choose instead an effective  linear confining potential $V = \lambda_F \zeta$ which reproduces the linear Regge behavior for baryons~\cite{Brodsky:2008pg, Abidin:2009hr}.
From  \req{Upm} we find for the effective potentials $U^\pm$ in Eqs. \req{LFWEA} and \req{LFWEB}
\beqa  \label{Uplus}
U^+(\zeta) &=& \lambda_F^2 \zeta^2 + 2 (\nu + 1) \lambda_F ,\\  \label{Uminus}
U^-(\zeta) &=& \lambda_F^2 \zeta^2 + 2 \nu  \lambda_F ,
\enqa
and the two-component solution
\beqa \label{psip}
\psi_+(\zeta) &\sim& 
  \zeta^{\frac{1}{2} + \nu} e^{-\vert \lambda_F\vert \zeta^2/2}
  L_{n}^{\nu}\left(\vert \la_F \vert \zeta^2\right),   \\ \label{psim}
\psi_-(\zeta) &\sim&  \zeta^{\frac{3}{2} + \nu} e^{-\vert \la_F \vert  \zeta^2/2}
 L_{n}^{\nu + 1}\left(\vert \la_F\vert \zeta^2\right).
\enqa

We can compute separately the eigenvalues for the wave equations \req{LFWEA} and \req{LFWEB} for arbitrary $\la_F$ and compare the results  for consistency, since the eigenvalues determined from both equations should be identical.  For the potential \req{Uplus} the eigenvalues of \req{LFWEA} are 
\beq \label{Mplus}
M_+^2 = \left(4n + 2 \nu + 2\right ) \vert \la_F  \vert + 2 \left( \nu  +1 \right) \la_F,
\enq
whereas for the potential \req{Uminus} the eigenvalues of \req{LFWEB} are
\beq
M_-^2 = \left(4n + 2(\nu +1) +2 \right) \vert \la_F \vert + 2 \nu \la_F.
\enq
For $\la_F>0$ we find $M_+^2 = M_-^2 = M^2$ where
\beq \label{M2F}
M^2 = 4 \, \lambda_F  \left(  n +  \nu + 1 \right),
\enq
identical for plus and minus eigenfunctions. For $\la_F<0$ it follows that $M_+^2 \ne M_-^2$ and no solution is possible. Thus the solution $\la_F <0$ is discarded.
Notice that in contrast with the meson spectrum \req{M2SFM} which depends on the quantum number  $J + L$, the baryon spectrum  \req{M2F} for $\nu = L+1$ and arbitrary $J$,  $M^2 = 4 \, \lambda_F  \left( n +  L + 2 \right)$, only depends on $L$, an important result also found in Ref. \cite{Gutsche:2011vb}.

It is important to notice that the solutions \req{psip} and \req{psim} of the second-order differential equations \req{LFWEA} and \req{LFWEB} are not independent
since the solutions must also obey the linear Dirac equation \req{LFDE} \cite{Mueck:1998iz}. This fixes the relative normalization. Using the relation 
$L_{n-1}^{\nu+1}(x) + L_n^\nu(x) = L_n^{\nu+1}(x)$ between the associated Laguerre functions we find for $\la_F >0$
\beqa
\psi_+(\zeta) &=& \la_F ^{(1+\nu)/2}\sqrt{\frac{2 n!}{(n+\nu - 1)!}} \,
  \zeta^{\frac{1}{2} + \nu} e^{-\lambda_F  \zeta^2/2}
  L_{n}^{\nu}\left(\la_F \zeta^2\right),   \\
\psi_-(\zeta) &=&  \la_F ^{(2+\nu)/2} \frac{1}{\sqrt{n + \nu + 1}} \sqrt{\frac{2 n!}{(n+\nu - 1)!}} \,
 \zeta^{\frac{3}{2} + \nu} e^{- \la_F   \zeta^2/2}
 L_{n}^{\nu + 1}\left(\la_F \zeta^2\right),
\enqa
with equal probability
\beq
\int d\zeta \, \psi_+^2(\zeta) =  \int d\zeta \, \psi_-^2(\zeta) =1.
\enq
If the plus solution represents  the $S$ component of a proton  and the minus  solution its $P$ component, it then follows that  the ``soft-wall''  holographic model for baryons discussed above is consistent with a  proton with $S$ and $P$ components with equal probability.  Consequently, its spin is carried out by the orbital angular momentum
$\langle J^z \rangle = \langle L^z \rangle = 1/2$, $\langle S^z\rangle = 0$, where $J^z = L^z + S^z$. Identical results follow for the hard-wall model of baryons.

Note that, as expected, the potential $\lambda_F^2 \zeta^2$ in the second order Dirac equations matches the soft-wall potential for mesons discussed in Sec.~\ref{LFmapM}, and thus we set  $\la_F = \la$  reproducing the universality of the Regge slope for mesons and baryons. However, the lowest possible eigenvalue for $n= L = 0$, the ground state in Eq. \req{M2F}, corresponds to the twist-2 trajectory $\nu = L$,  and not the twist-3 trajectory $\nu = L + 1$  determined by the short-distance scaling behavior. The twist-2 trajectory corresponds to an effective two-particle bound state, in this case the active quark versus the  spectators (a diquark) of the cluster decomposition from the holographic mapping.  Therefore \req{M2F}  does not give a good description of the Regge baryon intercepts.  This problem has been discussed in detail in Ref.  \cite{deTeramond:2012rt},  and the following relations have been inferred analytically. For the  positive-parity nucleon sector  
\begin{equation} 
M^{2 \, (+)}_{n, L, S} =  4 \la \left(n + L + \frac{S}{2} + \frac{3}{4} \right) \label{M2p},
\end{equation}
where the internal spin $S = \half$  or $\frac{3}{2}$. The corresponding formula for the negative-parity baryons is 
\begin{equation} 
M^{2 \,(-)}_{n, L, S} =  4 \la \left(n + L + \frac{S}{2} + \frac{5}{4} \right)  \label{M2m} ,
\end{equation}
with a mass gap  $2 \lambda$ for Regge trajectories with the same internal spin but opposite parity. 
Notice that  $M^{2 \,(+)}_{n, L, S = \frac{3}{2}} = M^{2 \,(-)}_{n, L, S =  \frac{1}{2}}$, and consequently the positive and negative-parity $\Delta$ states lie in the same trajectory, consistent with the experimental results.  

As discussed in~\cite{deTeramond:2012rt} the full baryon orbital and radial excitation spectrum is very well described by  \req{M2p} and \req{M2m}.  An important feature of light-front holography is that it predicts a similar multiplicity of states for mesons and baryons,  consistent with what is 
observed experimentally~\cite{Klempt:2007cp}. This remarkable property could have a simple explanation in the cluster decomposition of the
holographic variable \req{hologz}, which labels a system of partons as an active quark plus a system of $n-1$ spectators. From this perspective, a baryon with $n=3$ looks in light-front holography as a quark--scalar-diquark system. It is also interesting to notice that in the hard-wall model the proton mass is entirely due to the kinetic energy of the light quarks, whereas in the soft-wall model described here, half of the invariant mass squared $M^2$ of the proton is due to the kinetic energy of the partons, and half is due  to the confinement potential.

\section{Summary and Discussion \label{summ}}

Holographic QCD provides a remarkable first approximation to hadron physics based on the duality between  AdS space and light-front quantization in physical space-time. 
In this article we have derived hadronic  bound-state equations for particles with arbitrary spin starting from an effective invariant action in a higher dimensional classical gravitational theory. The fact that we can map the equations of motion from the gravitational theory to a Hamiltonian equation of motion in light-front quantized QCD has been our principal guide. The undisturbed AdS geometry reproduces the kinematical aspects of the light-front Hamiltonian, notably the emergence of a LF angular momentum
which is holographically identified with the mass in  the gravitational theory.    The breaking of the maximal symmetry of AdS  then allows the introduction of  the confinement dynamics  of the theory in physical space-time.

Thus in order to fully  preserve all the kinematical aspects, a consistent mapping to LF quantized QCD requires a clear separation between the kinematical and dynamical effects. The introduction of symmetry breaking effects in the action has to be carried out in such a way as to avoid interference between the two.
Although the kinematical aspects can be treated in parallel both for integer and half-integer spin states, the introduction of dynamics can be different for mesons and baryons.

In the approach discussed in this article for integer spin, confinement can be achieved  by imposing boundary conditions in the infrared region of AdS space, or by effectively modifying the infrared region of AdS by inserting a dilaton term in the effective action, or by explicitly distorting the metric of AdS space.  In addition, $z$-dependent AdS mass  terms are  introduced in the effective action which are uniquely determined  by the requirement of no mixing between kinematics  and dynamics. Following this procedure, one is led to a light-front potential which depends separately on the total angular momentum $J$ and the LF angular momentum $L$, and it agrees  with the light-front model of Ref.~\cite{deTeramond:2008ht} which describes the meson spectrum  very well~\cite{deTeramond:2012rt}.

The requirement to clearly separate  kinematical and dynamical aspects becomes especially evident for spins higher than 1. For spin-0, the covariant derivative coincides with the partial one, and for spin-1, the action can be constructed in such a way as to eliminate the affine connection (Appendix \ref{vect}). Thus no interference occurs in this case. 
For higher spins, however, one has to deal with higher-rank symmetric tensors, and therefore the contribution of the affine connection cannot be discarded. 
Furthermore, for higher spin states  many different ways of contracting the tensor indices of the spinor fields and the derivatives in the action are possible. These different contractions are necessary in order to obtain the subsidiary conditions required to eliminate the lower spin states. For higher spin the choice of the contractions becomes very complex and as a practical procedure, we  choose an effective action with a very simple contraction scheme, where the intricacies of the different contractions and mixing effects from dynamics are assumed to be absorbed in the $z$ dependence of an effective AdS mass term. Remarkably,  this simple choice yields for integer spin all the subsidiary conditions  necessary to eliminate the lower spin states in physical space-time.

In the case of  half-integer spin, our  effective action  leads to a Dirac-like equation which can be mapped to the LF Hamiltonian bound-state equation. This effective action  also leads to the Rarita-Schwinger condition for the spin index.  Since the action is linear in the covariant derivatives,  the contribution of the dilaton or an additional warping factor of the metric can be absorbed into a redefinition of the spinor fields, and no dynamical terms appear in the resulting equations of motion.
Therefore, the dilaton does not  lead to confinement~\cite{Kirsch:2006he}.  Nonetheless, one can obtain a discrete spectrum for baryons by introducing confinement either by imposing boundary conditions~\cite{deTeramond:2005su},    or  by an additional effective interaction in the Lagrangian~\cite{Brodsky:2008pg, Abidin:2009hr}.
Since no mixing occurs in this case,   no $z$-dependent mass terms in the AdS action are necessary.

We now turn to specific models. For the  hard-wall model the treatment of higher spin is very simple. The kinematics are fully reproduced by the invariant effective action \req{action2} without  explicit $z$-dependent symmetry-breaking terms. Since the dynamics is encoded exclusively in the boundary conditions, no mixing between dynamical and kinematical effects occurs, and consequently no $z$-dependent mass term is necessary.   This has as a consequence that  the resulting spectrum in the hard-wall model, does not depend on $J$ explicitly, but only on the light-front angular momentum $L$.

In contrast, the results of the soft-wall model for integer spin, either with a dilaton factor or with an additional warping factor of the metric, agree exactly with those of  Refs. \cite{deTeramond:2008ht, deTeramond:2012rt} and yield good agreement with the data. The sign of the dilaton profile $\vp(z)= \la z^2$ is uniquely fixed in our approach, namely $\la > 0$.   The solution $\la < 0$ is incompatible with the light-front constituent interpretation of hadronic bound states. In particular, the  solution $\la >0$ gives a massless pion, consistent with the zero quark mass chiral limit of QCD~\cite{Brodsky:2013yy}.
In contrast the negative sign dilaton  leads to a pion mass larger than that of the $\rho$-meson.

On the other hand, the approach of Ref. \cite{Karch:2006pv} requires a negative dilaton profile, $\la <0$, in order to obtain a rising vector meson trajectory.  This approach 
is based on different assumptions: it starts from a gauge-invariant theory in AdS. In a specific gauge, no terms from the affine connection appear in the action, and the AdS mass has to be fixed to be zero in that gauge. Since there is no freedom in the choice of the AdS mass, it is not possible to introduce the light-front orbital angular momentum of the constituents independently of $J$.  Therefore, this approach is incompatible with the mapping of the AdS equations of motion to the light-front Hamiltonian for bound states.

For baryons the many-body state is described by an effective two-body light-front Hamiltonian, where the holographic variable is mapped to the invariant separation of one constituent (the active constituent) to the cluster of the rest (the spectators).  Therefore, the mapping of AdS equations to the light-front bound state equations predicts that there is only one  relevant angular momentum,  the light-front orbital angular momentum $L$ between the active and the spectator cluster. Furthermore, since the action for fermions is linear in the covariant derivatives, no mixing between dynamical and kinematical aspects occurs. Thus, for fermions there is no explicit $J$ dependence in the light-front equations of motion, and thus the bound-state spectrum of baryons can only depend on $L$.

These remarkable predictions, which are inferred from the geometry of AdS space, are independent of the specific mechanisms of symmetry breaking and account for many the striking similarities and  differences observed in the systematics of the meson and baryon spectra.    The equality of the slopes of the Regge trajectories and the multiplicity of states for mesons and baryons is explained. We also explain the  observed differences in the meson versus baryon spectra that are due to spin-orbit coupling.  For example, the predicted triplet spin-orbit splitting for vector mesons  is in striking contrast with the empirical near-degeneracy of baryon states of different total angular momentum $J$; the baryons are  classified by the internal orbital angular momentum quantum number $L$  along a given Regge trajectory, not $J.$
There are, however, other remarkable regularities in the baryon trajectories, which can be inferred from the data~\cite{deTeramond:2012rt}, but are not deduced systematically from the AdS effective action.  In particular, the Regge intercepts of the baryon trajectories are not consistent with the data. This open problem indicates that there are still essential elements missing in the description of baryons in light-front holographic QCD.

\acknowledgments

This work  was supported by the Department of Energy contract DE--AC02--76SF00515.

\appendix

\section{Integer Spin in AdS Space\label{intspinA}}

We label $x^M = \left(x^\mu, z\right)$, with $M, N = 0, \dots , d$, the coordinates of AdS$_{d+1}$ space  and 
 $\mu, \nu= 0, 1, \dots, d-1$ the Minkowski flat space-time indices. The AdS metric tensor  in Poincar\'e's coordinates is
 \beq  \label{ads=metric}  
 g_{MN} = \frac{R^2}{z^2}\, \eta_{MN} ,
 \enq
where $\eta_{MN}$ is the flat $d + 1$ metric  $(1, -1, \cdots, - 1)$.
The corresponding vielbein follows from $g_{M N} = e^A_M e^B_N  \eta_{A B}$  and is given by
\beq
 e^A_M =
 \frac{R}{z}  \de^A_M ,
\enq 
where  $A, B = 0, \dots , d$  are tangent AdS space indices and the flat metric  $\eta_{A B}$ has diagonal components  $(1, -1, \cdots, -1)$.   
To simplify the notation we shall use in the  appendices the following convention for the indices: 
$\{N\}=\{N_1 N_2 \cdots N_J\}$ and
${\{LN_{/j}\}}= \{L \,N_1 \cdots N_{j-1} N_{j+1} \cdots N_J\}$. \,
Furthermore, we define \,
$g^{\{N N'\}} = g^{N_1N_1'} \cdots g^{N_J N_J'}$.

\subsection{Covariant Derivatives for Integer Spin \label{der-int}}

We compute the covariant derivatives using the affine connection for the AdS metric given by the Christoffel symbols
\beqa \lb{gammads}
\Gamma^{L}_{MN} &=& \frac{1}{2}  g^{LK}\left(\pa_M g_{KN} +\pa_N
 g_{KM}- \pa_K  g_{MN}\right)\nn \\
  &=& -\Omega(z) \left(\de^z_M\de^L_N+\de^z_N\de^L_M -
  \eta^{Lz}\eta_{MN}\right) ,
  \enqa
with the warp factor $\Omega(z) = 1/z$ in AdS space.  
We find
 \beqa D_M \Phi_{\{N\}} &=& \pa_M \Phi_{\{N\}}-\sum_j
\Ga^L_{MN_j} \Phi_{\{LN_{/j}\}} \nn \\
&=&  \pa_M \Phi_{\{N\}} + \Omega(z)  \sum_j \Big( \de^z_M
\Phi_{\{N_j N_{/j}\}}+\de^z_{N_j} \Phi_{\{MN_{/j}\}} +
\eta_{M N_j} \Phi_{\{z N_{/j}\}}\Big) ,
\enqa
and thus
\beqa  D_z \Phi_{\{N\}}&=&
 \pa_z \Phi_{\{N\}} + \Omega(z)  \sum_j \Big( \de^z_M
\Phi_{\{N_j N_{/j}\}}+\de^z_{N_j} \Phi_{\{zN_{/j}\}} +
\eta_{z N_j} \Phi_{\{z N_{/j}\}}\Big)  \nn \\
&=& \pa_z \Phi_{\{N\}}+ J \Omega(z)  \, \Phi_{\{N\}}  , \\
  D_\mu \Phi_{\{N\}}
  &=& \pa_\mu \Phi_{\{N\}} + \Omega(z)  \sum_j \Big( \de^z_{N_j} \Phi_{\{\mu N_{/j}\}} +
\eta_{\mu N_j} \Phi_{\{z N_{/j}\}}\Big).
 \enqa

It is convenient to work with coordinates in the local tangent frame
 \beq \label{inert}
 \hat \Phi_{\{A\}} = e^{\{N\}}_{\{A\}}  \Phi_{\{N\}}=
 \left(\frac{z}{R}\right)^J  \Phi_{\{A\}},
\enq
where we find
\beq \label{dz}
  D_z \Phi_{\{N\}}= \left(\frac{R}{z}\right)^J \pa_z \hat \Phi_{\{N\}},
\enq
 and
\beq \label{dmu} 
 g^{\mu \mu'}   g^{\{\nu \nu'\}}
D_\mu \Phi_{\{\nu\}} \, D_{\mu'} \Phi_{\{\nu'\}}= g^{\mu \mu'} \eta^{\{\nu \nu'\}}
\left( \pa_\mu \hat \Phi_{\{\nu\}} \, \pa_{\mu'} \hat \Phi_{\{\nu'\}} +  g^{zz}
J \Omega^2(z) \, \hat \Phi_{\{\nu\}} \, \hat \Phi_{\{\nu'\}}\right) .
\enq

 \subsection{Spin-1 Vector  Field in AdS Space \label{vect}}

To illustrate the effect of the different contractions for the tensor fields in the equations of motion discussed in Sec. \ref{int},
we derive in this section the equations of motion for a vector field.  We start with the generalized Proca-action for a vector field in AdS$_{d+1}$ space
\beq  \label{SV}
S =  \int \! d^d x \, dz  \,\sqrt{\vert g \vert} \,e^{\varphi(z)}
  \left( \frac{1}{4} g^{M R} g^{N S} F_{M N} F_{R S} - \half \, \mu^2 g^{M N} \Phi_M \Phi_N \right),
 \enq
where $F_{MN} = \partial_M \Phi_N - \partial_N \Phi_M$.  The variation of the action  leads  to  the equation of motion
 \beq \label{VAdS}
 \frac{1}{ \sqrt{g} \, e^\varphi} \partial_M \left(\sqrt{g} \, e^\varphi g^{M R} g^{N S} F_{R S} \right) + \mu^2 g^{N R} \Phi_R = 0,
  \enq
together with the supplementary condition 
\beq \label{Lorentz}
\partial_M \left( \sqrt{g} e^\varphi g^{M N} A_N \right) = 0.
\enq

Using the AdS metric (\ref{ads=metric}) and the condition (\ref{Lorentz}) we can express  (\ref{VAdS}) as a system of coupled differential  equations
\beqa\label{WEV1}
\left[  \eta^{\mu \nu}  \partial_\mu \partial_\nu  
-  \frac{ z^{d-1}}{e^{\varphi(z)}}   \partial_z \left(\frac{e^{\varphi(z)}}{z^{d-1}} \partial_z\right)  
  - \partial_z^2 \varphi
+  \left(\frac{\mu R}{z}\right)^2 + 1 -d\right] \Phi_z &\! = \!& 0 , \\ \label{WEV2}
 \left[  \eta^{\mu \nu} \partial_\mu \partial_\nu 
- \frac{ z^{d-3}}{e^{\varphi(z)}}   \partial_z \left(\frac{e^{\varphi(z)}}{z^{d-3}} \partial_z\right) 
+ \left(\frac{\mu R}{z}\right)^2\right] \Phi_\mu & \! =  \! & -  \frac{2}{z} \partial_\mu \Phi_z .
\enqa

In the physical subspace defined  by $\Phi_z =0$  the system of coupled differential equations (\ref{WEV1}\,-\ref{WEV2}) reduces to 
\beq 
\left[  \eta^{\mu \nu} \partial_\mu \partial_\nu 
- \frac{ z^{d-3}}{e^{\varphi(z)}}   \partial_z \left(\frac{e^{\varphi(z)}}{z^{d-3}} \partial_z\right) 
+ \left(\frac{\mu R}{z}\right)^2\right] \Phi_\mu   =  0.
\enq
Thus, the constant AdS mass $\mu$ appearing in the full action (\ref{SV})  is also the mass in the covariant equation of motion,  and no further $z$-dependent AdS mass shift  is necessary to separate the kinematical and dynamical components. In this case, the antisymmetric contraction has eliminated the contribution from the affine connection and no interference between kinematical and dynamical effects occurs.

\subsection{\label{app-el} Separation of Kinematical and Dynamical Aspects in the Equations of Motion}

In this appendix we show that the $z$ dependence of the effective mass $\mu_{\it eff}$ in the effective action $S_{\it eff}$ \req{action2} is determined by the
distinct separation of kinematical and dynamical aspects. As emphasized in this article,  kinematical effects are determined 
by the AdS geometry and the dynamical effects are caused by the breaking of the maximal symmetry in the action, {\it e.g.},  by introducing a dilaton. In order 
to isolate the kinematical terms we separate in the action \req{action2}  the contributions of the affine connections into a distinct term $P[\Phi]$
\beq \label{action2pt}
 S_{\it eff} =  \int d^{d} x  \, dz \, \sqrt{|g|}  \, e^{\vp(z)} g^{\{ N N' \}} 
 \Big(g^{M M'} \pa_M \Phi^*_{\{N\}} \, \pa_{M'} \Phi_{ \{N ' \}}  
 - \mu_{\it eff}^2(z) \Phi^*_{\{N\} }\, \Phi_{\{N ' \}}\Big)+ P[\Phi]  . 
\enq

Purely kinematical effects from  the affine connection are absent in the equations of motion derived from  $S_{\it eff} -P[\Phi]$.
The influence of dynamics can be eliminated by setting  $\vph(z) =0$.
Since $ S_{\it eff} -P[\Phi]$ contains only partial derivatives, the Euler-Lagrange equations for  this truncated action, which contains no contribution from the affine connection, are easily obtained
\beq 
 \left[ - \frac{z^{d-1-2J}}{e^{\vp(z)}}\pa_z\left( \frac{e^{\vp(z)}}{z^{d-1-2J} }\right)+
 \frac{ (\mu_{\it eff}(z) R)^2 }{z^2}  \right]  \Phi_J = \cM^2 \, \Phi_J.
 \label{eomnoncov} 
 \enq
On the other hand, the equations of motion derived from the full action \req{action2} are given by (\ref{PhiJM}) and (\ref{muphi})
\beq 
 \left[ - \frac{z^{d-1-2J}}{e^{\vp(z)}}\pa_z\left( \frac{e^{\vp(z)}}{z^{d-1-2J} }\right) +
 \frac{( \mu_{\it eff}(z) R)^2 - J\,z\,\vp'(z) + J(d-J+1)}{z^2}  \right] \Phi_J = \cM^2 \, \Phi_J.
 \label{eomcov} \enq

The difference between  \req{eomnoncov} and \req{eomcov} shows that  the affine connection only contributes to an AdS mass-like term. Part of the difference  is independent of the dilaton $\vp(z)$, {\it i.e.},  kinematical.   This constant term is however not essential, since the constant contribution to the AdS mass is not an {\em a priori} determined parameter, but determined by the light-front angular momentum $L$.
There is however a term in the difference, which is proportional to $\vp'(z)$,
{\it  i.~e.},  it is  due to an  interference between the dynamics and kinematics. To keep the separation between the kinematical and dynamical effects,
this term has to be compensated by an appropriate choice of the $z$ dependence of the effective mass $\mu_{\it eff}$ in \req{eomcov} 
\beq  \label{muel} 
(\mu_{\it eff}(z)R)^2=  J \,z\, \vp'(z) + C,
\enq
where $C$ is a constant.  Setting  $ C = m^2 - J(d - J + 1)$ we recover (\ref{muphiz}).

In the case where the maximal symmetry of the AdS metric is not
broken by a dilaton,  $\vp(z)=0$, no $z$-dependent mass shift is necessary and one can start with a constant mass in \req{action2}.
This is the case in the hard-wall model, where  the dynamical effects are introduced by  the boundary conditions and indeed no mixing between kinematical and dynamical aspects does occur.

 \subsection{Warped Metric \label{wrapmet}}

In this Appendix we investigate the effects of conformal symmetry breaking starting with the warped metric (\ref{gw}) with metric tensor and vielbein
 \beq \tilde
g_{MN}=\frac{R^2}{z^2} e^{2 \tilde \vp(z)} \, \eta_{MN},  \quad   \quad
 \tilde e^A_M =
 \frac{R}{z} e^{\tilde \vp(z)} \de^A_M ,
\enq
and no dilaton background. The Christoffel symbols for  the warped metric (\ref{gw}) have the same form as (\ref{gammads}) with the warp factor
$\tilde \Omega(z) = 1/z  - \pa_z \tilde \vp(z)$.

 The effective action is
\beq
\label{action1w} 
 \tilde S_{\it eff} = \int d^{d} x \, dz \sqrt{\vert \tilde g \vert} \,  \tilde g^{\{N N'\}} \left(  \tilde g^{M M'} 
 D_M \Phi^*_{\{N\}}\, D_{M'} \Phi_{\{N'\}} - \tilde \mu_{\it eff}^2(z) \Phi^*_{\{N\}}\, \Phi_{\{N'\}}\right),
\enq
where $\tilde \mu_{\it eff}(z)$ is the effective mass.

We can express the covariant derivatives in (\ref{action1w})
  in terms of partial derivatives in a local tangent frame
 \beq \label{inert}
 \hat \Phi_{\{A\}} = e^{\{N\}}_{\{A\}}  \Phi_{\{N\}}=
 \left(\frac{z}{R}\right)^J  \, e^{-J\,\tilde \vp(z)} \, \Phi_{\{A\}}.
\enq
We obtain
\beq 
\label{dz}
  D_z \Phi_{\{N\}}= \left(\frac{R}{z}\right)^J e^{ J \tilde \vp(z)}\pa_z \hat 
\Phi_{\{N\}} ,
\enq
 and
\beq \label{dmu} \tilde g^{\mu \mu'}  \tilde g^{\{\nu \nu'\}}
D_\mu \Phi_{\{\nu\}} \, D_{\mu'} \Phi_{\{\nu'\}}
= \tilde g^{\mu \mu'} \eta^{\{\nu \nu'\}}
\left( \pa_\mu \hat \Phi_{\{\nu\}} \,
\pa_{\mu'} \hat \Phi_{\{\nu'\}} 
+ \tilde g^{zz} \, J \, \tilde \Omega^2(z) \, \hat \Phi_{\{\nu\}} \, \hat \Phi_{\{\nu'\}}\right) .
\enq

Following exactly the same steps as described in Sec.  \ref{int} lead now to
\begin{multline}
 \label{a-wraphw}
\tilde S_{\it eff}^{[0]}  = \int d^{d} x \, dz \left(\frac{R \, e^{\tilde \vp(z)}}{z}\right)^{d-1}  \!
 \eta^{\{\nu \nu'\}} \Bigg( - \pa_z \hat \Phi^*_{\{\nu\}}\,\pa_z \hat \Phi_{\{\nu'\}}
  + \eta^{\mu \mu'}
\pa _\mu \hat \Phi^*_{\{\nu\}}\,\pa_{\mu'} \hat \Phi_{\{\nu'\}} \\
- \left[   \left(\frac{\mu_{\it eff}(z) R \, e^{\tilde \vp(z) }}{z} \right)^2 + J  \tilde \Omega^2(z)\right]
 \hat \Phi^*_{\{\nu\}}\, \hat \Phi_{\{\nu'\}}\Bigg).
\end{multline}

Comparing (\ref{a-wraphw}) with the AdS action (\ref{a-wraph}), we see that both forms of the action are equivalent
provided that  we set
\beq \label{subw} 
\tilde \vp(z) ={\textstyle \frac{1}{d-1}} \vp(z)
\quad \mbox{and} \quad 
(\tilde \mu_{\it eff}(z) R)^2 e^{2 \tilde \vp}  =(\mu_{\it eff}(z) R)^2 + \tilde \Omega^2(z) (J -1).
\enq
Thus, the warp-metric action $\tilde S_{\it eff}^{[0]}$ agrees with the dilaton action
$S_{\it eff}^{[0]}$, Eq. (\ref{a-wraph}), leading  to the same results, notably the bound state equations \req{PhiJM}
from which we obtain the relation
\beq
\left(\tilde \mu_{\it eff}(z) R \right)^2 = \left(m^2 + J z \frac{\tilde \vp '(z)}{d-1} - J z^2 \tilde \Omega^2(z) - J(d-J) \right) e^{- 2 \tilde \vp(z)}
\enq

For the Euler Lagrange equations derived from $\tilde S^{[1]}_{\it eff}$, the term in the warped action equivalent to (\ref{a-subh}),  the warp factor $\tilde \Om$ factors out and 
its special form is therefore not relevant for the kinematical conditions derived from \req{ELz}. We therefore obtain the same kinematical constraints which eliminates the lower spin states as for the dilaton case discussed in Sec.~\ref{int}.

\section{Half-Integer Spin in AdS Space\label{intspinB}}

Using the notation of Appendix \ref{intspinA} we write the covariant derivative of a  Rarita-Schwinger spinor $\Psi_{\{N\}}$
\beq
 D_M \Psi_{\{N\}} = \pa_M \Psi_{\{N\}} 
 - \frac{i}{2} \om_M^{AB} \Si_{AB} \Psi_{\{N\}}  - \, \sum_j \Ga ^L_{M N_j} \Psi_{\{L N_{/j}\}} ,
\enq
where $\Sigma_{A B}$ are
the generators of the Lorentz group in the spinor representation
\beq
\Sigma_{A B} = \frac{i}{4}  \left[\Gamma_A, \Gamma_B\right] ,
\enq
and the tangent space Dirac matrices obey the usual anti-commutation relation 
 \beq \label{Ga} \Ga ^A \, \Ga^B + \Ga^B\, \Ga^A = 2 \, \eta^{AB}. \enq
 The spin connection in AdS is 
 \beq  \label{afc}
 w_M^{A B} = \Om(z) \left(\eta^{A z} \delta_M^B - \eta^{B z} \delta^A_M\right), 
 \enq
with  $\Om(z) = 1/z$ and the Christoffel symbols are defined in Appendix \ref{intspinA}.
 
 For  even $d$ we can choose the set of gamma matrices
$\Gamma^A = \left(\Gamma^\mu, \Gamma^z\right)$ with 
$\Gamma^z =  \Gamma^0 \Gamma^1 \cdots \Gamma^{d-1}$.
For $d=4$ one has
\beq \label{gamma4}
 \Ga^\mu = \ga^\mu, \quad \Ga^z =  - \Gamma_z=  -i \, \ga^5 ,
\enq
 where $\ga^\mu$ and  $\ga^5$
are the usual 4-dimensional Dirac matrices with $\gamma^5 \equiv i \gamma^0 \gamma^1 \gamma^2 \gamma^3$ and $(\ga^5)^2=+1$.
The spin connections are given by 
 \beq \label{spinconads}
  \om^{z \al}_\mu = - \om^{\al z}_\mu = \Om(z) \de^\al_\mu ,
  \enq
  all other components $\om^{A B}_M$  are zero.

The covariant derivatives of a Rarita-Schwinger spinor in AdS are
\beqa  D_z \Psi_{\{N\}}
&=& \pa_z \Psi_{\{N\}}+ T\, \Om(z) \Psi_{\{N\}} 
= \left(\frac{R}{z}\right)^T \pa_z \hat  \Psi_{\{N\}} ,\\
  D_\mu \Psi_{\{N\}}&=&
 \pa_\mu \Psi_{\{N\}} + \frac{1}{2 } \Om(z) \Ga_\mu \, \Ga_z  \Psi_{\{N\}}  +
 \Om(z) \sum_j \Big( \de^z_{N_j} \Psi_{\{\mu N_{/j}\}} +
\eta_{\mu N_j} \Psi_{\{z N_{/j}\}}\Big) .
\nn
 \enqa
From these equations one obtains easily \req{afnu} and \req{afz}.

\subsection{Spin-$\frac{3}{2}$ Rarita-Schwinger Field in AdS Space \label{threehalf}}

The generalization~\cite{Volovich:1998tj, Matlock:1999fy} of the Rarita-Schwinger action~\cite{Rarita:1941mf}  to 
AdS$_{d+1}$  is
\beq
S = \int
d^{d} x \,dz\,  \sqrt{| g|}\; \bar \Psi_N \left( i \tilde \Ga^{[NMN']}\,D_M -\mu \, \tilde \Ga^{[NN']} \right)\Psi_N , \enq
where $\tilde \Ga^{[NMN']}$ and $\tilde \Ga^{[NN']}$ are the antisymmetrized products of three and two Dirac matrices $\tilde \Ga^M = e^M_A \Ga^A = \frac{z}{R} \, \de^M_A\,\Ga^A$, with tangent space matrices $\Ga^A$ given by \req{Ga}.
From the variation of this action  one obtains  the generalization of the Rarita-Schwinger equation
\beq \label{aS}
 \left(i\,\tilde \Ga^{[NMN']}\,D_M -\mu\, \tilde  \Ga^{[NN']}\right)\Psi_{N'} = 0 .
 \enq
 The Christoffel symbols in the covariant derivative can  be omitted  due to the  the antisymmetry of the indices in $\tilde \Ga^{[NMN']}$ and only the spin connection must be taken into account. Eq.  \req{aS}  leads to  the Rarita-Schwinger condition~\cite{Volovich:1998tj}
 \beq
\Ga^M\, \Psi_M=0 ,\enq
and the generalized Dirac equation~\cite{Matlock:1999fy}  
\beq \label{PsihatB}
\left[ i \left( z \eta^{M N} \Gamma_M \partial_N + \frac{d}{2} \Gamma_z \right) - \mu R \right]   \hat \Psi_A = \Ga_A \hat \Psi_z ,
\enq
for the spinor with tangent indices $\hat \Psi_A =   \frac{z}{R} \, \de^M_A\,  \Psi_M$.
These equations agree for $T=1,\;  \vp(z)=\rho(z) =0$  and  $\hat \Psi_z=0$ with  Eq. \req{Psihat}, derived from the effective action \req{af}.

\end{document}